\documentclass[aps, prd, twocolumn, superscriptaddress, nofootinbib, floatfix, noshowpacs]{revtex4-2}
\usepackage{amsmath,placeins,graphicx,amssymb,tabularx}
\allowdisplaybreaks  
\usepackage[percent]{overpic}
\usepackage{color,xcolor}
\definecolor{prdblue}{rgb}{0.133,0.118,0.498}
\usepackage[colorlinks=true, pdfstartview=FitV, linkcolor=prdblue, citecolor= prdblue, urlcolor=prdblue]{hyperref}
\usepackage{orcidlink} 
\usepackage{slashed}

\begin{document}
\title{Physics-informed neural networks for quark Dyson--Schwinger equations and propagator singularities}

\newcommand{\orcidauthorA}{\orcidlink{0009-0008-1946-1997}} 
\newcommand{\orcidauthorB}{\orcidlink{0009-0004-5306-147X}} 

\author{Xin-Peng Li \orcidauthorA{}}
\email[]{xpli@smail.nju.edu.cn}
\affiliation{School of Physics, \href{https://ror.org/01rxvg760}{Nanjing University}, Nanjing, Jiangsu 210093, China}
\affiliation{Institute for Nonperturbative Physics, \href{https://ror.org/01rxvg760}{Nanjing University}, Nanjing, Jiangsu 210093, China}
\affiliation{European Centre for Theoretical Studies in Nuclear Physics and Related Areas (\href{https://ror.org/01gzye136}{ECT*}), Villa Tambosi, Strada delle Tabarelle 286, I-38123 Villazzano (TN), Italy}

\author{Zhong-Hua Huang \orcidauthorB{}}
\affiliation{School of Software, \href{https://ror.org/023hj5876}{Dalian University of Technology}, Dalian, Liaoning 116620, China}

\date{\today}

\begin{abstract}
We develop a physics-informed neural-network (PINN) framework for solving the rainbow-truncated quark Dyson--Schwinger equation at real and complex momenta. The resulting dressing functions agree closely with conventional iterative solutions. We investigate the singularity structure of the quark propagator using two complementary methods. Analytic continuation of a PINN trained on the real momentum axis estimates the locations of poles and further reveals three complex-conjugate pole pairs. An independent reconstruction using PINN-parametrized spectral densities and a single pole pair yields $q=-0.146\pm0.391i~\mathrm{GeV}^2$. The agreement between the two methods substantially strengthens confidence in the properties of the leading singularities. Both spectral densities change sign, and the negative regions of the vector density and the complex poles indicate a violation of reflection positivity, consistent with quark confinement. Finally, we use the extracted poles and residues to calculate their contributions to two order parameters at finite quark chemical potential, the chiral condensate and the quark number density.
\end{abstract}

\maketitle

\section{\label{Introduction}Introduction}
Quantum chromodynamics (QCD) provides the fundamental description of strong interactions. Although perturbation theory is highly effective at large momentum scales owing to asymptotic freedom, the growth of the interaction strength in the infrared renders perturbative methods inadequate~\cite{Binosi:2016nme,Cui:2019dwv,Deur:2022msf,Deur:2023dzc}. Understanding infrared phenomena such as dynamical chiral symmetry breaking, confinement, and hadron bound states therefore requires nonperturbative approaches.

Continuum Schwinger function methods (CSMs) are one approach to nonperturbative QCD~\cite{Xu:2026mds,Zeng:2026xsr,Bai:2026nqo,Binosi:2026mej,Bender:1996bm}. The Dyson--Schwinger equations (DSEs) form an infinite hierarchy of coupled integral equations for the Green's functions of the theory, and the DSE for the dressed-quark propagator is known as the quark gap equation~\cite{Binosi:2025kpz,Yao:2025fnb,Binosi:2026tre,Roberts:1992jy,Roberts:1994dr}. In practice, the truncated equations are commonly solved using iterative numerical methods on discrete momentum grids. Although highly effective, these methods naturally produce discrete solutions, so subsequent calculations of hadron spectra and other observables may require interpolation or additional reconstruction procedures.

Hadron bound-state calculations require the quark propagator at complex momenta, and the accessible bound-state mass range is limited by its nonanalytic structures~\cite{Xu:2025cyj,Xu:2024nzp,Yao:2024drm,Binosi:2023won}. Determining the propagator reliably in the complex plane and identifying its leading singularities are therefore important both for solving the Bethe--Salpeter equations and for understanding the analytic manifestations of quark confinement.

Originally developed for partial differential equations, physics-informed neural networks (PINNs) and related methods have been applied to fluid dynamics, plasma physics, and quantum many-body problems~\cite{MAO2020112789,Qin:2022hmd,Wu:2025wvv}. Recent applications include hadronic physics~\cite{Dai:2026tjl,Kaidisch:2023lwp}. PINNs represent solutions as continuous, differentiable functions and incorporate the governing equations into the training loss~\cite{Raissi:2017zsi, Raissi:2017gan}.

A single physics-informed neural network is used in Ref.~\cite{Terin:2024iyy} to solve a DSE for the dynamical mass function in quantum electrodynamics. Extending this approach to the quark gap equation is more challenging because one must resolve coupled dressing functions whose infrared and ultraviolet behaviors differ markedly in magnitude and variation scale.

In this work, we first employ PINNs to solve the quark DSE both on the real momentum axis and within a parabolic domain of the complex momentum plane. To determine the nonanalytic structure of the quark propagator, we use two complementary methods: analytic continuation of the real-axis solution and spectral reconstruction. We compare the position and residues of the leading pole pair obtained with the two methods. We also examine how these nonanalytic structures contribute to the chiral condensate and the quark number density, two order parameters used in effective-potential studies of the QCD phase structure at finite quark chemical potential~\cite{Gao:2025kzk}.

This paper is organized as follows. We begin in Sec.~\ref{Dyson-Schwinger Equations} with the framework used in this work. We obtain PINN solutions for the $N^+$, $N^-$, and Wigner branches on the real momentum axis and in parabolic domains of the complex momentum plane, as presented in Sec.~\ref{sec:QP}. In Sec.~\ref{sec:SS}, the singularity structure of the vacuum $N^+$ propagator is investigated using two complementary methods, analytic continuation and spectral reconstruction, whose predictions for the leading pole pair are compared. We also examine how the poles and the continuous spectrum contribute to the chiral condensate and quark number density at zero temperature and finite chemical potential. Our findings and prospects for future work are summarized in Sec.~\ref{sec:summary}, while details of the network architectures, momentum sampling, and training loss are given in Appendix~\ref{Appendix A}.

\section{\label{Dyson-Schwinger Equations}Quark gap equation and interaction model}
The gap equation is the natural starting point for analyses of dynamical chiral symmetry breaking (DCSB) and confinement~\cite{Binosi:2022djx,Roberts:1994dr,Roberts:2000aa,Krassnigg:2004jz,Li:2025cll}. It also underpins the formulation of a symmetry-preserving, Poincar\'e-covariant treatment of bound states via the Bethe--Salpeter and Faddeev equations~\cite{Qin:2018dqp,Xiao:2025cqz,Eichmann:2009en}. In Euclidean space, the gap equation takes the form
\begin{align}
    S&^{-1}(p)=Z_2i\gamma\cdot p+Z_4m(\zeta)\nonumber\\
    +&Z_1\int_q^\Lambda g^2D_{\mu\nu}(p-q)\frac{\lambda^a}{2}\gamma_\mu S(q)\frac{\lambda^a}{2}\Gamma_\nu(q,p), 
\end{align}
where $D_{\mu\nu}$ is the gluon propagator, $\Gamma_\nu$ is the quark--gluon vertex, $\int_q^\Lambda$ denotes a Poincar\'e-invariant regularization of the four-dimensional integral, $m(\zeta)$ is the renormalized current-quark mass, and $Z_1$ and $Z_2$ are the quark--gluon vertex and quark wave-function renormalization constants, respectively. The coefficient of the mass term is $Z_4=Z_2 Z_m$, where $Z_m$ is the quark mass renormalization constant. These constants depend on the renormalization point $\zeta$ and the regulator scale $\Lambda$. The dressed-quark propagator can be written as
\begin{align}
S(p)=\frac{1}{i\gamma\cdot pA(p^2)+B(p^2)},
\end{align}
where $A(p^2)$ and $B(p^2)$ are the vector and scalar dressing functions, respectively. $M_q(p^2)=\frac{B(p^2)}{A(p^2)}$ is the dressed-quark mass function, which is renormalization group invariant in a multiplicatively renormalizable theory.

The gap equation belongs to an infinite tower of coupled DSEs involving higher $n$-point functions. Since solving the complete hierarchy is intractable, a symmetry-preserving truncation is required to obtain a closed and computable system.

A particularly successful and widely used symmetry-preserving truncation~\cite{Munczek:1994zz,Bender:1996bb} is the rainbow truncation $\Gamma_\nu(q,p)=\gamma_\nu$, which represents the leading-order term in a controlled, symmetry-consistent expansion. 

Studies of the gauge sector in QCD motivate an efficacious parametrization for the product of the effective charge and the gluon two-point Schwinger function~\cite{Yao:2024ixu,Qin:2011dd,Binosi:2014aea,Binosi:2016wcx},
\begin{eqnarray}
    Z_1g^2D_{\mu\nu}(k) = Z_2^2\mathcal{G}(k^2)T_{\mu\nu}(k),
\end{eqnarray}
where $k^2T_{\mu\nu}(k)=k^2\delta_{\mu\nu}-k_\mu k_\nu$, and
\begin{eqnarray}
\mathcal{G}(k^2)=\frac{8\pi^2}{\omega_0^4}De^{-k^2/\omega_0^2}+\frac{8\pi^2\gamma_m}{\ln\left[\tau+(1+k^2/\Lambda_{QCD}^2)^2\right]}\mathcal{F}(k^2),
\end{eqnarray}
where $\gamma_m=12/(33-2N_f)$, $N_f=4$, $\Lambda_{QCD}=0.234~\mathrm{GeV}$; $\tau=e^2-1$; and $\mathcal{F}(s)=\{1-\exp{(-s/[4m_t^2])}\}/s$, $m_t=0.5~\mathrm{GeV}$. This form of the interaction preserves the one-loop renormalization-group behavior of QCD within the gap equation and has been widely employed in the successful prediction and interpretation of a broad range of hadronic observables~\cite{Zeng:2026peb,Xiao:2025cqz,Yao:2025xjx}. Herein, the parameter set $D=1.024~\mathrm{GeV^2}$, $(\omega_0 D)^{\frac{1}{3}}=0.8~\mathrm{GeV}$, $m(\zeta) =0.0034~\mathrm{GeV}$ is chosen to reproduce the pion mass $M_\pi=0.134~\mathrm{GeV}$ and decay constant and also provides a good description of light-hadron observables~\cite{Qin:2018dqp}.

\section{Quark propagator at real and complex momenta}\label{sec:QP}

To solve the quark gap equation accurately, we construct separate PINNs for the real momentum axis and for parabolic domains in the complex momentum plane. The resulting quark dressing functions provide the numerical basis for our analysis of the propagator's nonanalytic structure in Sec.~\ref{sec:SS}.

\subsection{Solutions on the real momentum axis}\label{subsec:SR}

After performing the traces over Dirac and color spaces, we obtain the following equations for $A(p^2)$ and $B(p^2)$:
\begin{subequations}\label{AB}
    \begin{align}
        A(p^2)=&Z_2+\frac{4Z_2^2}{3p^2}\int_q^\Lambda\mathcal{G}((p-q)^2)\frac{A(q^2)}{q^2A^2(q^2)+B^2(q^2)}\nonumber\\
        &\times(p\cdot q+2\frac{((p-q)\cdot q)((p-q)\cdot p)}{(p-q)^2}),\\
        B(p^2)=&Z_4m(\zeta)+4Z_2^2\int_q^\Lambda\mathcal{G}((p-q)^2)\frac{B(q^2)}{q^2A^2(q^2)+B^2(q^2)}.
    \end{align}
\end{subequations}
Denoting the integral contributions to the vector and scalar equations by $\Sigma_A(p^2,\Lambda^2)$ and $\Sigma_B(p^2,\Lambda^2)$, respectively, one has
\begin{subequations}
\begin{align}
A(p^2)&=Z_2+\Sigma_A(p^2,\Lambda^2),\\
B(p^2)&=Z_4m(\zeta)+\Sigma_B(p^2,\Lambda^2).
\end{align}
\end{subequations}
At the renormalization point $\zeta=19~\mathrm{GeV}$~\cite{Chang:2008ec}, we impose the momentum-subtraction conditions
\begin{subequations}
\begin{align}
A(\zeta^2)&=1,\\
B(\zeta^2)&=m(\zeta).
\end{align}
\end{subequations}
The renormalization constants are therefore determined by
\begin{subequations}
\begin{align}
Z_2&=1-\Sigma_A(\zeta^2,\Lambda^2),\\
Z_4&=1-\frac{\Sigma_B(\zeta^2,\Lambda^2)}{m(\zeta)}.
\end{align}
\end{subequations}
Substituting these expressions into the coupled equations for $A$ and $B$ yields their renormalized form
\begin{subequations}
\begin{align}
A(p^2)&=1+\Sigma_A(p^2,\Lambda^2)-\Sigma_A(\zeta^2,\Lambda^2),\\
B(p^2)&=m(\zeta)+\Sigma_B(p^2,\Lambda^2)-\Sigma_B(\zeta^2,\Lambda^2).
\end{align}
\end{subequations}

\begin{figure}[!tbp]
\centering
\begin{overpic}[width=\columnwidth]{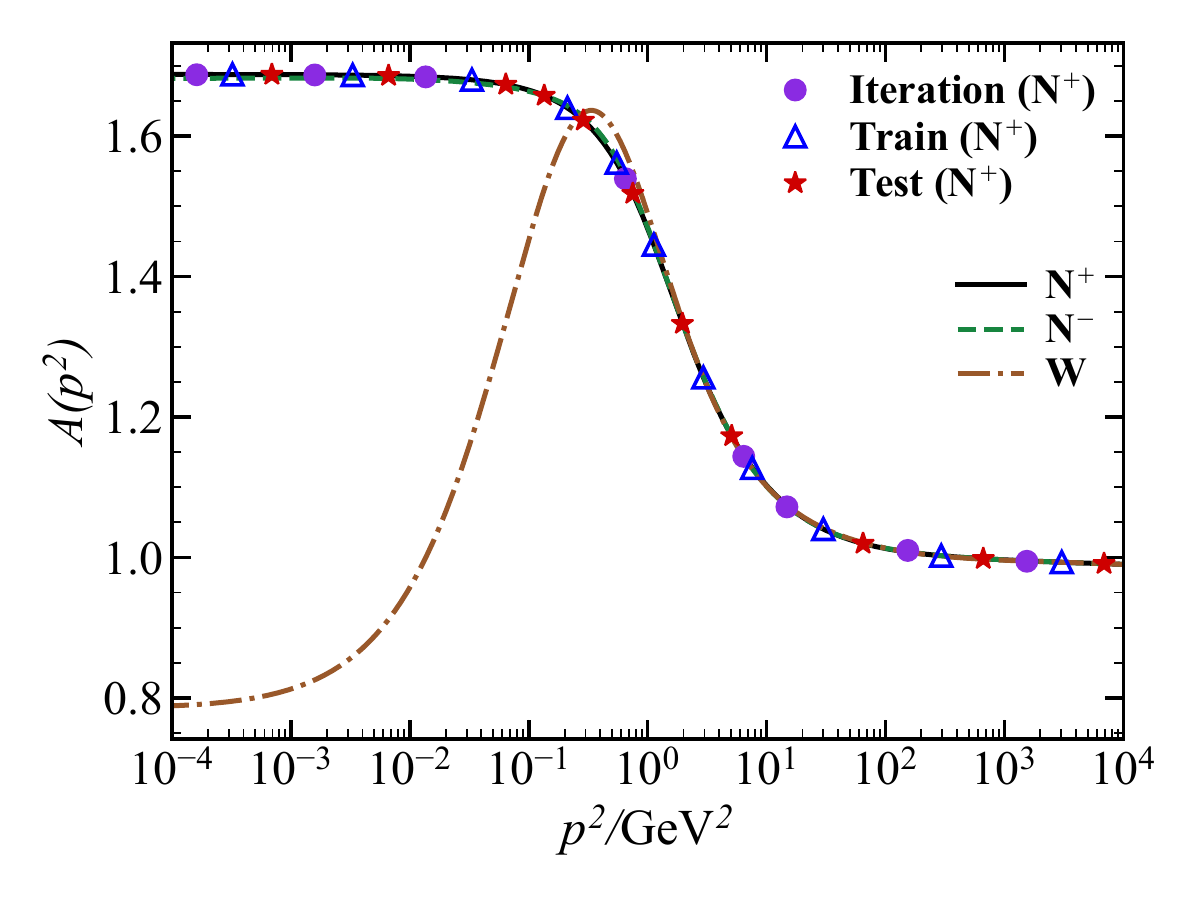}
    \put(2,65){\textsf{\large A}}
\end{overpic}
\\
\begin{overpic}[width=\columnwidth]{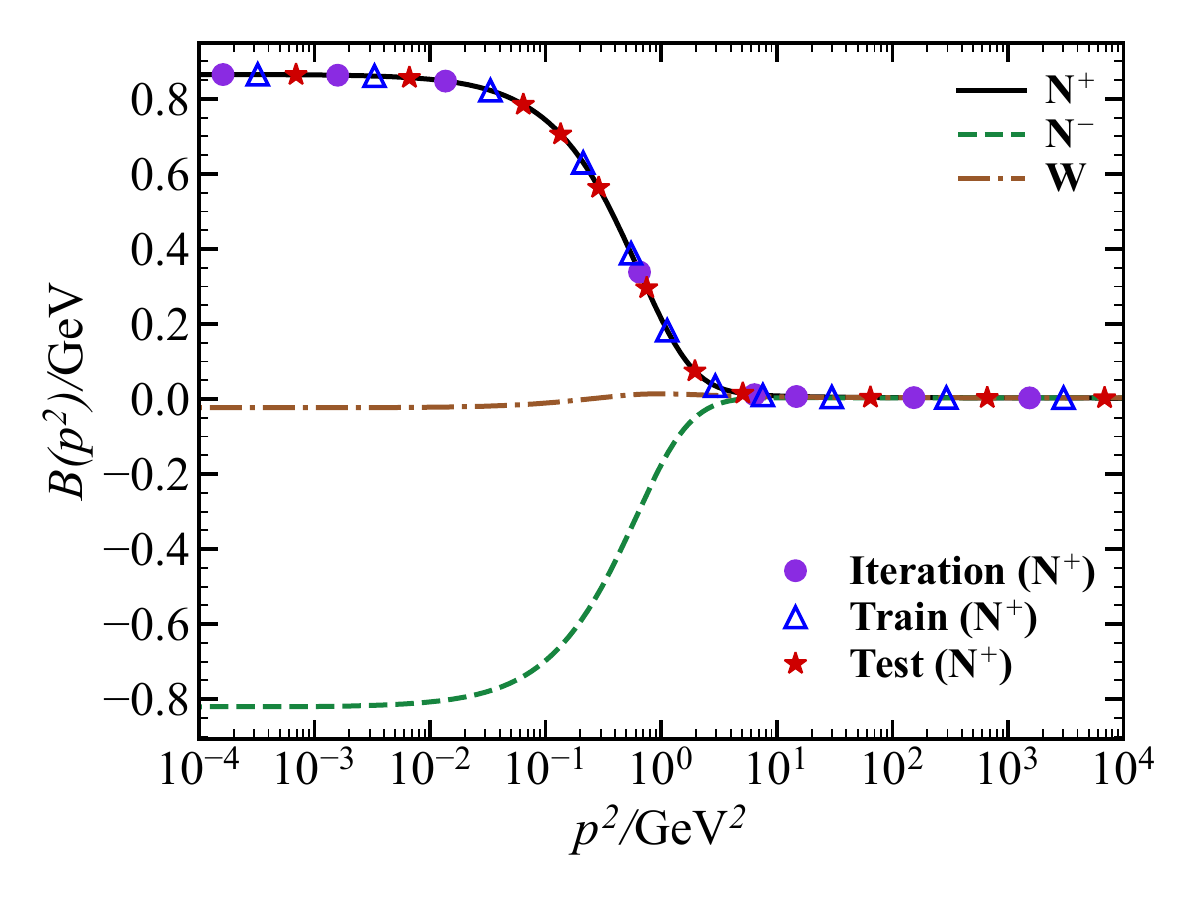}
    \put(2,65){\textsf{\large B}}
\end{overpic}
\\
\begin{overpic}[width=\columnwidth]{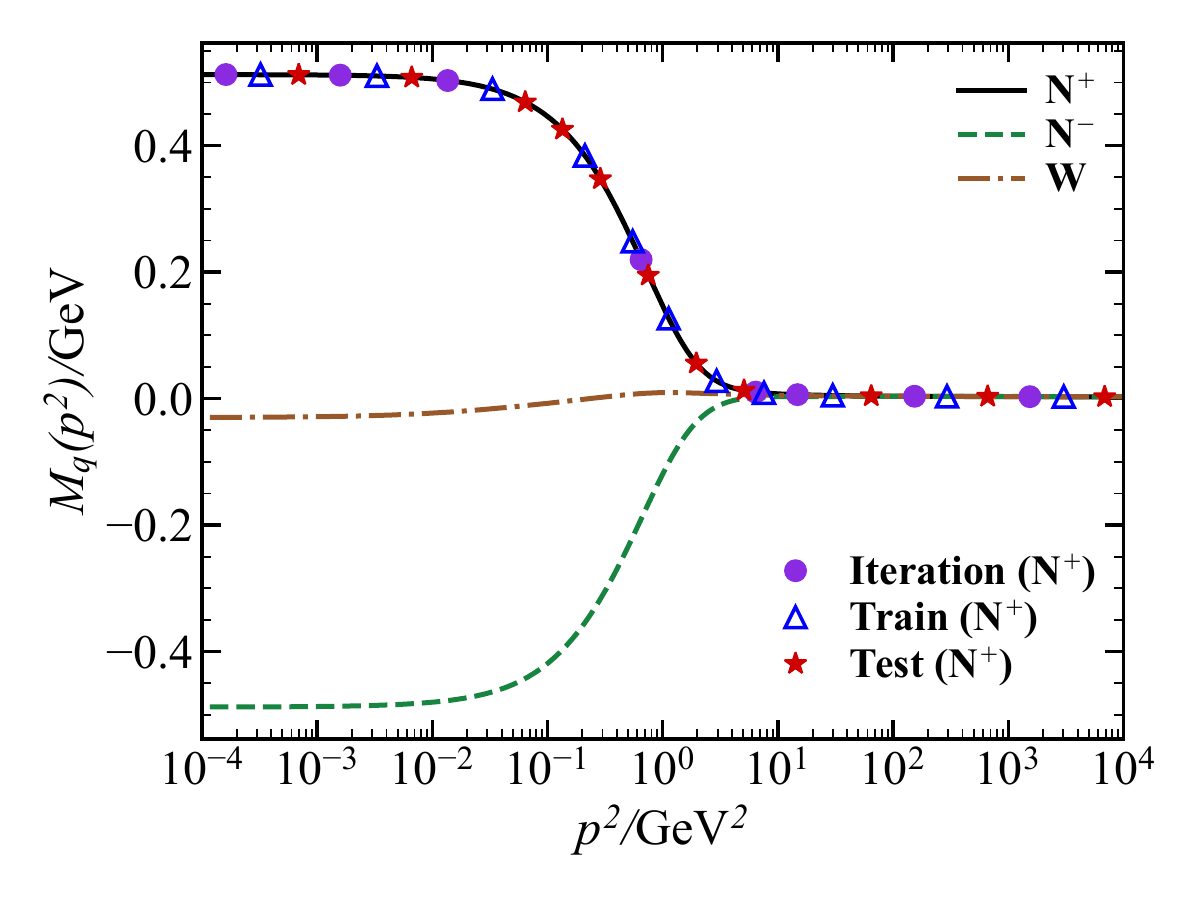}
    \put(2,65){\textsf{\large C}}
\end{overpic}
\caption{Three coexisting solutions of the quark gap equation on the real momentum axis.
\textsf{Panel A.} Vector dressing function $A(p^2)$.
\textsf{Panel B.} Scalar dressing function $B(p^2)$.
\textsf{Panel C.} Quark mass function $M_q(p^2)$. $M_q^{N^+}(0)=0.52~\mathrm{GeV}$.
\textsf{Curves.} In all three panels, the solid black, dashed green, and dash-dotted brown curves show the continuous PINN solutions for the positive Nambu ($N^+$), negative Nambu ($N^-$), and Wigner ($W$) branches, respectively.
\textsf{Points.} All markers refer to the $N^+$ branch. Purple circles show the conventional iterative solution. Open blue triangles and red stars show the PINN data at selected training and test points, respectively.}
\label{fig1}
\end{figure}

Solving these equations, we obtain numerical results shown in the black line of Fig.~\ref{fig1}. We sample 270 distinct points on the real momentum axis using two Gauss--Legendre grids, one for training (open blue triangles, 120 points, 44.4\% of the total) and the other for testing (red stars, 150 points, 55.6\% of the total). The solutions from the conventional iterative method and the PINN framework are in excellent agreement with each other, with mean relative errors of $0.0281\%$ for $A(p^2)$, $2.38\%$ for $B(p^2)$, and $2.36\%$ for $M_q(p^2)$. Meanwhile, the test-set performance of the PINN remains satisfactory without obvious overfitting behavior. The consistent prediction across different methods and the reliable generalization capability demonstrate that the PINN-based solver is valid and accurate enough for practical DSE calculations.

As shown in Fig.~\ref{fig3}A, the training and test losses for the solution on the real momentum axis follow similar trends. One epoch corresponds to a complete pass through the training grid, during which the gap equation residual is evaluated at all training points and the network parameters are updated in a single optimization step. The loss decreases rapidly over the first 6000 epochs and then more gradually over the remaining 14,000 epochs.

\begin{figure*}[!tp]
\centering
\begin{overpic}[width=0.48\textwidth]{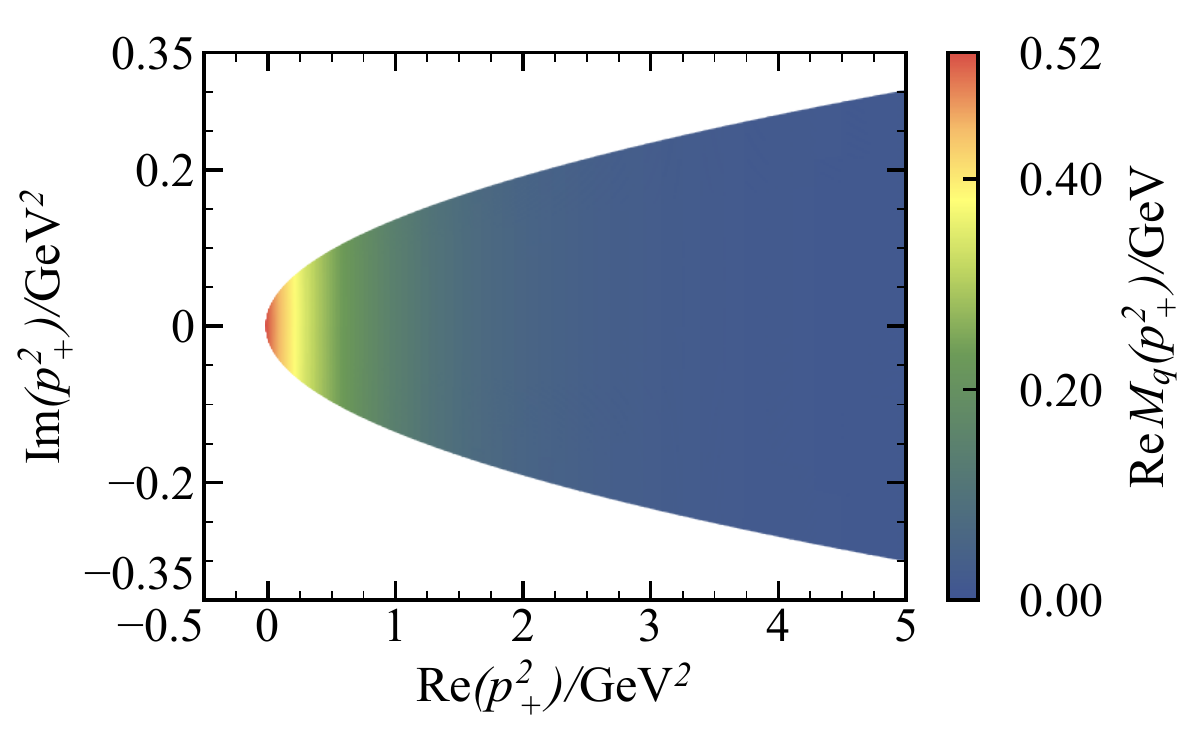}
    \put(2,56){\textsf{\large A}}
\end{overpic}
\hfill
\begin{overpic}[width=0.48\textwidth]{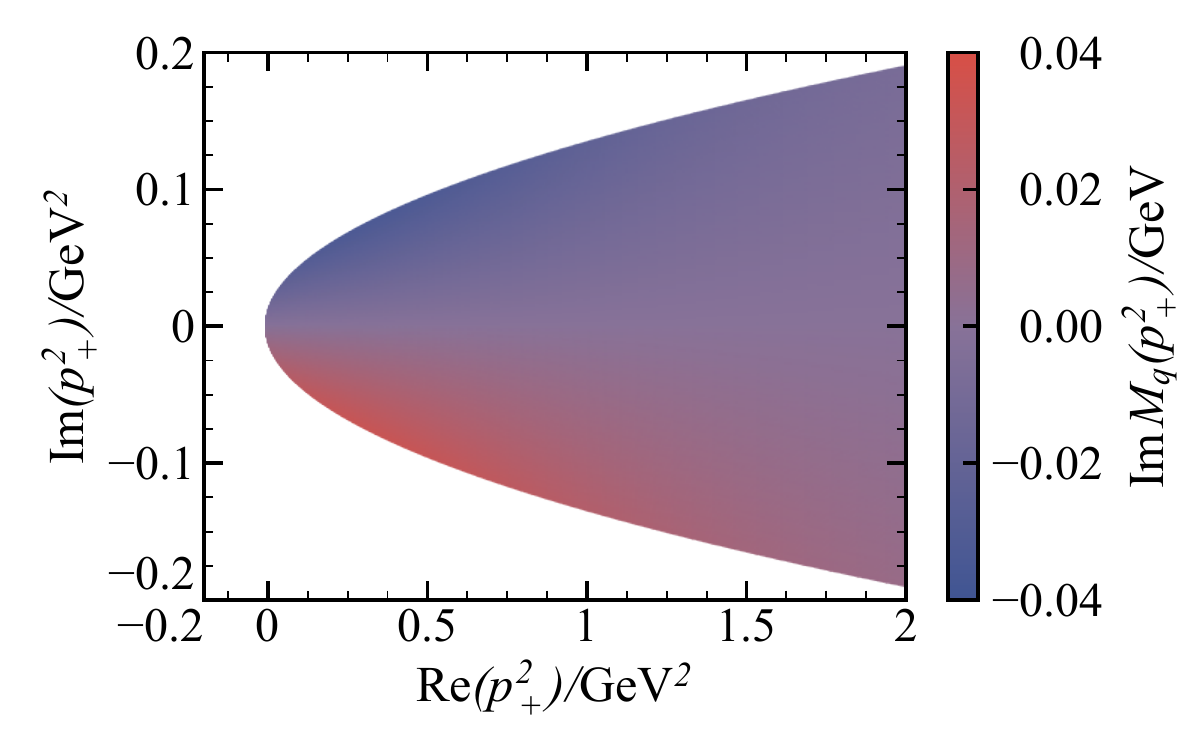}
    \put(2,56){\textsf{\large B}}
\end{overpic}\\
\begin{overpic}[width=0.48\textwidth]{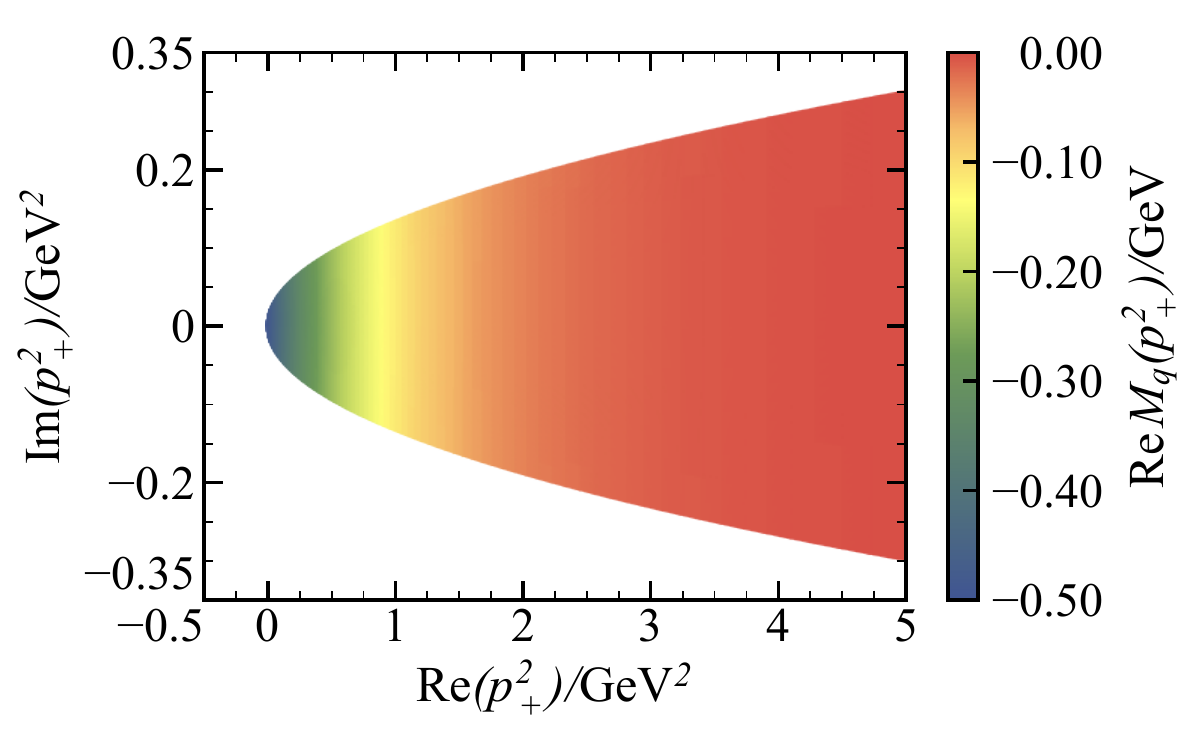}
    \put(2,56){\textsf{\large C}}
\end{overpic}
\hfill
\begin{overpic}[width=0.48\textwidth]{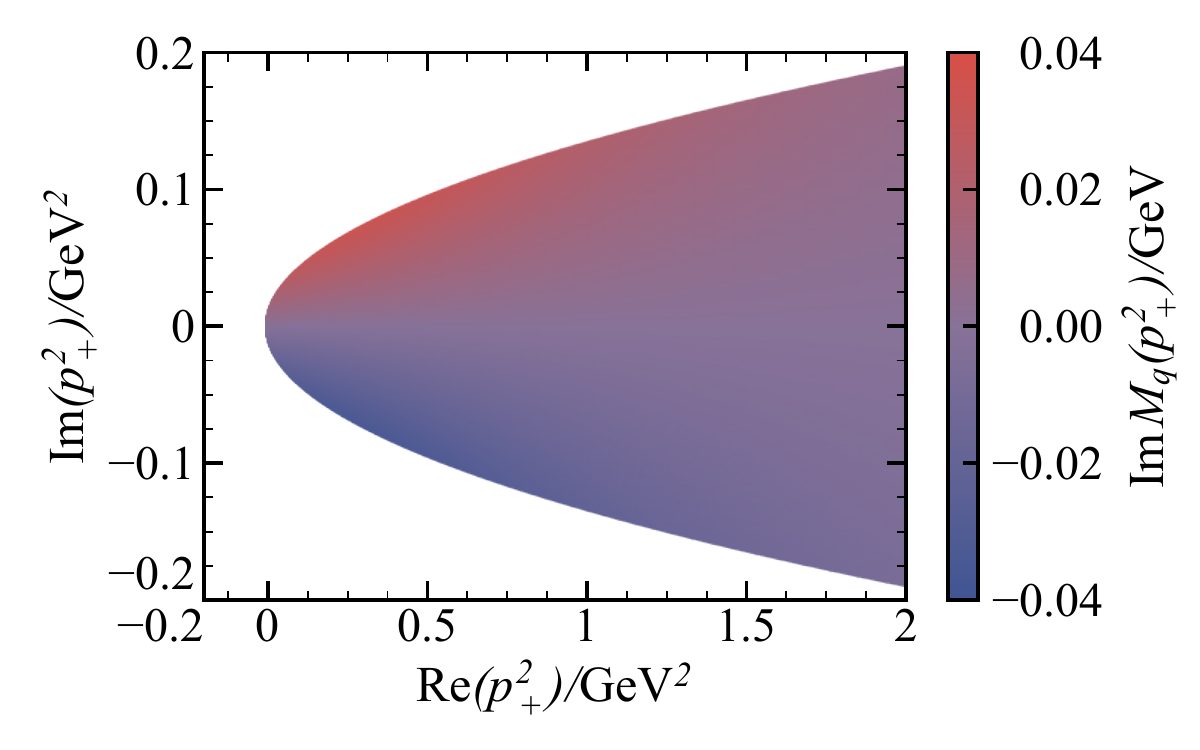}
    \put(2,56){\textsf{\large D}}
\end{overpic}\\
\begin{overpic}[width=0.48\textwidth]{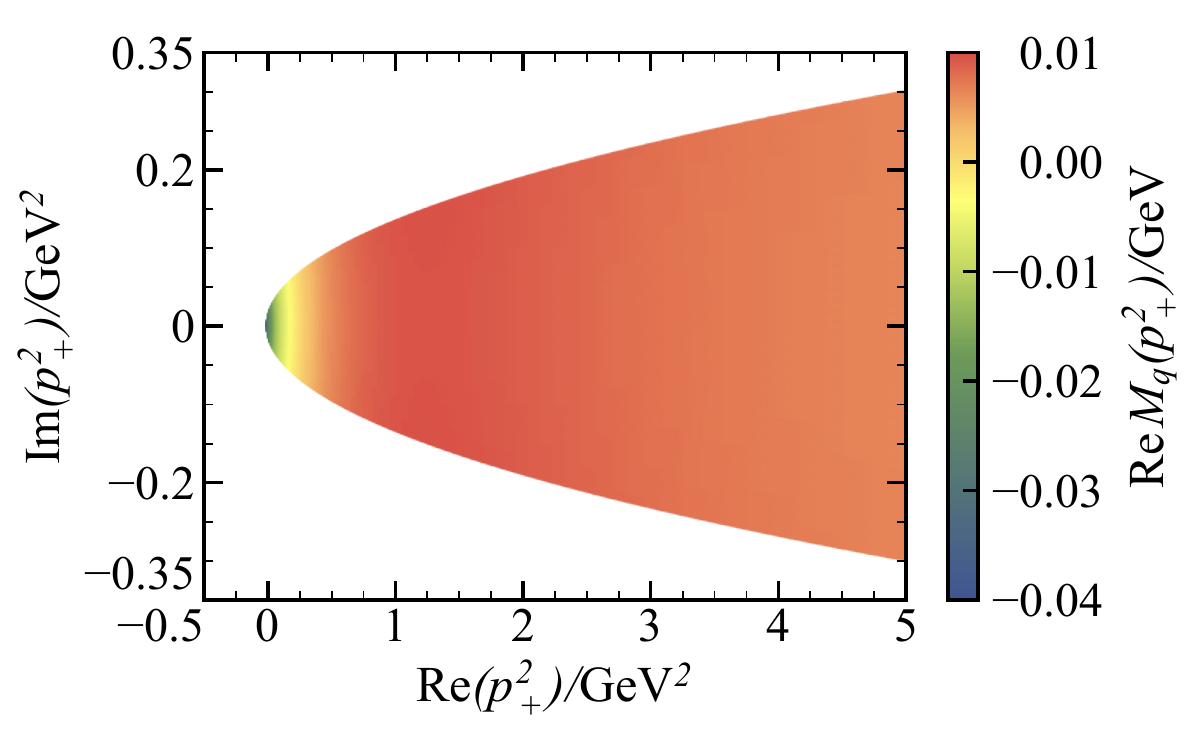}
    \put(2,56){\textsf{\large E}}
\end{overpic}
\hfill
\begin{overpic}[width=0.48\textwidth]{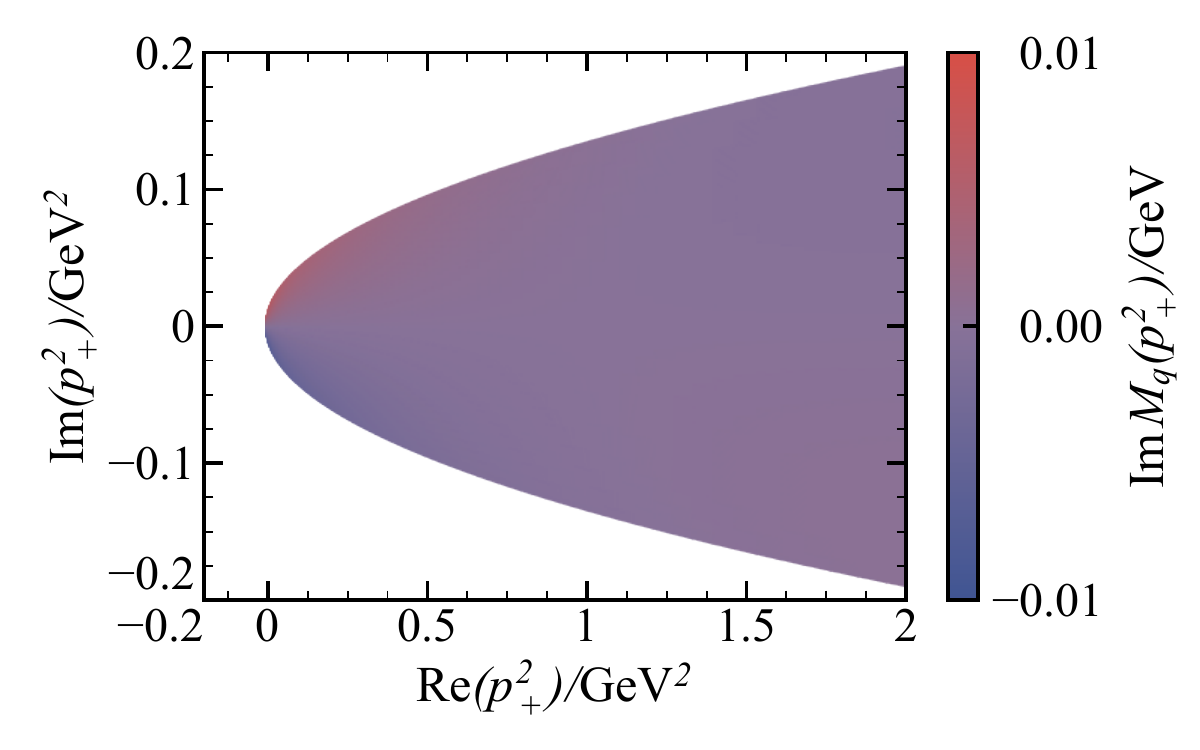}
    \put(2,56){\textsf{\large F}}
\end{overpic}
\caption{PINN solutions for the quark mass function $M_q(p_+^2)=B(p_+^2)/A(p_+^2)$ at pion mass $M=0.134~\mathrm{GeV}$ and $\eta=0.5$. \textsf{Panels A, B.} Positive Nambu branch $N^+$. \textsf{Panels C, D.} Negative Nambu branch $N^-$. \textsf{Panels E, F.} Wigner branch $W$. From high to low values, the color scales run from red through orange, yellow, and green to blue for the real part (left), and from red through purple to blue for the imaginary part (right).
\label{fig2}}
\end{figure*}

For the parameter set used here, the quark gap equation, Eq.~(\ref{AB}), admits three distinct solution branches, conventionally denoted by $N^+$ (positive Nambu), $N^-$ (negative Nambu), and $W$ (Wigner). In the chiral limit $m(\zeta)=0~\mathrm{GeV}$, both Nambu branches exhibit dynamical chiral symmetry breaking. Since the gap equation is invariant under $B(p^2)\to -B(p^2)$ at fixed $A(p^2)$, the two branches have nonzero mass functions of equal magnitude and opposite sign, $M_q^{N^-}(p^2)=-M_q^{N^+}(p^2)$. The Wigner solution preserves chiral symmetry and has $M_q^W(p^2)\equiv 0$. The nonzero current-quark mass used here explicitly breaks this symmetry, lifting the exact sign relation between the Nambu branches and allowing a nonzero mass function on the Wigner branch~\cite{Wang:2012me,Li:2025cll}. Our PINN solver is capable of computing all three solution branches with high accuracy, as shown in Fig.~\ref{fig1}. In contrast to the conventional iterative method, which requires careful tuning of convergence directions to target a specific branch, the PINN converges to a designated solution simply by using branch-specific initializations shown in Appendix~\ref{Output layer}.

\subsection{Solutions in parabolic domains of the complex momentum plane}\label{subsec:SC}

Our calculations are performed in Euclidean space, where the total four-momentum $P$ of a meson of mass $M$ satisfies the on-shell condition $P^2=-M^2$. In the meson rest frame, the fourth component of $P$ is therefore purely imaginary. Since the quark and antiquark each carry a share of this total momentum, their squared momenta generally take complex values even when the relative momentum is real. Solving the meson Bethe--Salpeter equation (BSE) thus requires the quark propagator in the complex momentum plane.

\begin{figure}[!bp]
\centering
\begin{overpic}[width=\linewidth]{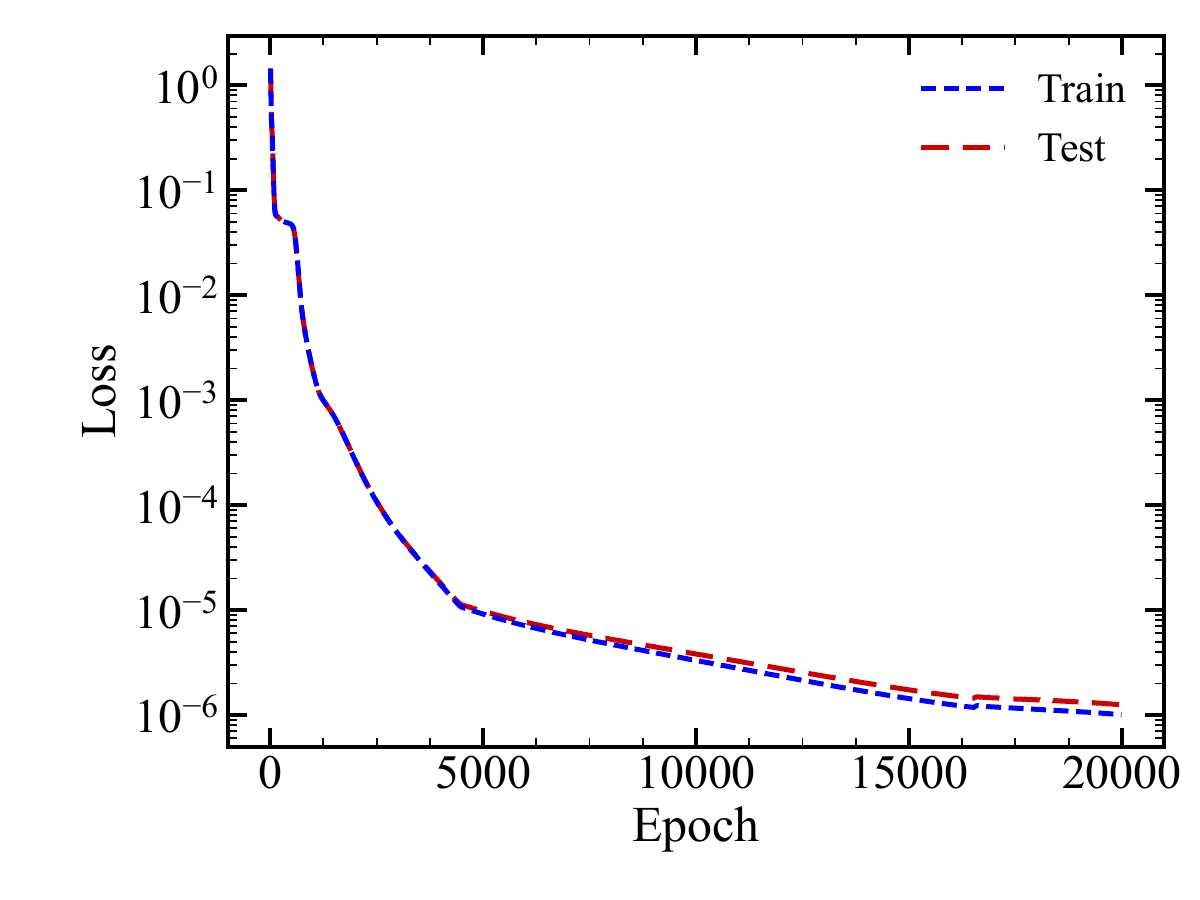}
    \put(2,58){\textsf{\large A}}
\end{overpic}
\\
\begin{overpic}[width=\linewidth]{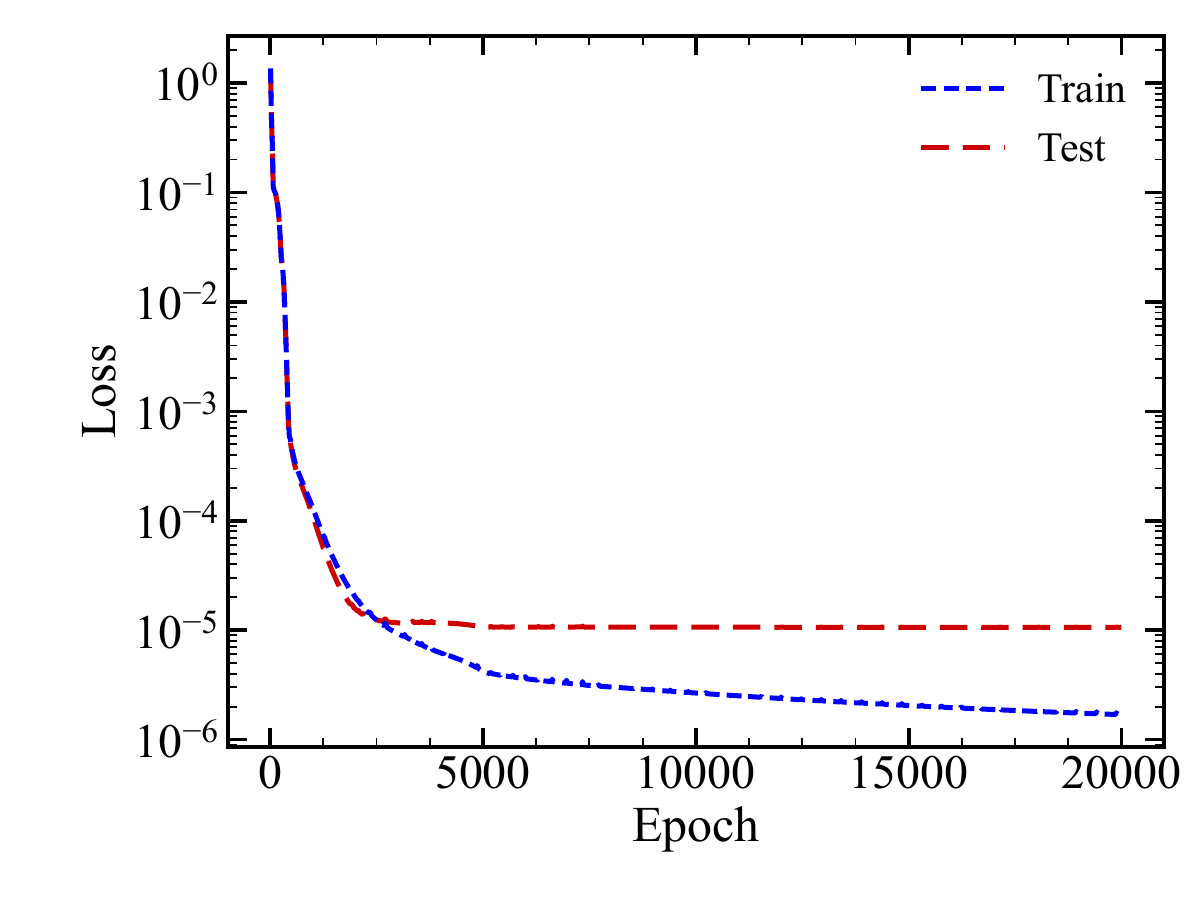}
    \put(2,58){\textsf{\large B}}
\end{overpic}
\caption{\textsf{Panel A.} PINN loss versus training epoch for the solution on the real momentum axis shown in Fig.~\ref{fig1}. Both the training and test losses reach the $10^{-6}$ level. \textsf{Panel B.} Corresponding losses for the solution in the complex momentum plane shown in Fig.~\ref{fig2}. The training loss also reaches the $10^{-6}$ level, while the test loss levels off near $10^{-5}$.
\label{fig3}}
\end{figure}

Analytic continuation and a shift of the integration momentum~\cite{Fischer:2005en} yield the coupled equations for $A(p_+^2)$ and $B(p_+^2)$,
\begin{subequations}
    \begin{align}
        A(p_+^2)=&Z_2+\frac{4Z_2^2}{3p_+^2}\int_q^\Lambda\mathcal{G}((p-q)^2)\frac{A(q_+^2)}{q_+^2A^2(q_+^2)+B^2(q_+^2)}\nonumber\\
        &\times(p_+\cdot q_++2\frac{((p-q)\cdot q_+)((p-q)\cdot p_+)}{(p-q)^2}),\\
        B(p_+^2)=&Z_4m(\zeta)+4Z_2^2\int_q^\Lambda\mathcal{G}((p-q)^2)\frac{B(q_+^2)}{q_+^2A^2(q_+^2)+B^2(q_+^2)}.
    \end{align}
\end{subequations}
Here, $p_+=p+\eta P$ and $q_+=q+\eta P$, where $p$ and $q$ are the real external and loop momenta, respectively. The momentum-partitioning parameter $\eta=0.5$ specifies the fraction of the meson's total four-momentum $P$ carried by the quark.

Once the dressing functions have been determined on a closed contour $\mathcal C$, their values at an interior point $z_0$ can be reconstructed using Cauchy's integral formula,
\begin{align}\label{cauchy}
f(z_0)=\frac{1}{2\pi i}
\oint_{\mathcal C}\frac{f(z)}{z-z_0}\,dz.
\end{align}
Hence, knowledge of the dressing functions on the parabolic contour permits their reconstruction throughout the enclosed domain, provided that the dressing functions are analytic on and inside the contour. The corresponding parabolic domain is $\operatorname{Re}(p_+^2)\geq\frac{[\operatorname{Im}(p_+^2)]^2}{4\eta^2M^2}-\eta^2M^2$.

Solving these equations yields the results shown in Fig.~\ref{fig2}. We sample 2460 distinct points in the complex momentum plane using two $n_p\times n_z$ grids with $n_p$ radial nodes and $n_z$ angular nodes: a $40\times24$ training grid with 960 points (39.0\%) and a $50\times30$ test grid with 1500 points (61.0\%). The mean relative differences between the PINN and iterative results are $0.0630\%$ for $A$, $6.08\%$ for $B$, and $6.07\%$ for the mass function $M_q$, respectively, demonstrating the high overall accuracy of the PINN solutions across the sampled complex-momentum domain.

As shown in Fig.~\ref{fig3}B, both losses decrease rapidly during the first 3000 epochs. The training loss then decreases from approximately $9.3\times10^{-6}$ to $1.7\times10^{-6}$, while the test loss levels off near $1.06\times10^{-5}$ and shows no appreciable increase. This behavior provides no clear evidence of overfitting during the later stages of training.

Finally, the three solution branches found on the real momentum axis persist in the complex momentum plane, where each branch can be understood as the analytic continuation of its real-axis counterpart. Fig.~\ref{fig2} shows the real and imaginary parts of the mass function $M_q$ for each branch: panels A and B for $N^+$, C and D for $N^-$, and E and F for the $W$ branch. The mean relative difference between the PINN and reference solutions for the mass function $M_q$ is $4.12\%$ on the $N^-$ branch and $1.50\%$ on the Wigner branch. This agreement shows that the PINN reproduces good overall accuracy across the sampled domain.

\section{Singularity structure of the quark propagator}\label{sec:SS}

We investigate the positions and residues of the dominant poles of the quark propagator on the $N^+$ branch. We first characterize the leading singularities of the quark propagator by analytically continuing the solution from the real momentum axis into the complex plane. Then, a spectral representation of the quark propagator is constructed independently by jointly fitting continuous vector and scalar spectral densities and a single complex-conjugate pole pair to the DSE solutions. The two approaches provide a mutual cross-check, and their agreement substantially strengthens confidence in the properties of the leading singularities. Finally, we use these results to evaluate the pole and spectral contributions to typical order parameters at zero temperature and finite quark chemical potential.

\subsection{Pole positions and residues from analytic continuation}\label{subsec:PP}

We compare the analytic continuation of the real-axis solution with the solution computed directly in the complex momentum plane. The direct calculation evaluates the quark gap equation at complex external momenta. If the dressing functions are analytic in a connected domain containing an interval of the real momentum axis, their values on that interval uniquely determine their continuation throughout the domain.

This observation provides an independent way to examine the reliability of the direct numerical solution at complex momenta. When a singularity or branch cut enters the parabolic domain, the analyticity required for reconstruction from the contour may be lost, leading to an instability or systematic deviation of the resulting solution. We therefore train an analytic PINN by imposing the gap equation only at real momenta and continue the learned dressing functions into the complex momentum plane. Its predictions are compared with the dressing functions obtained by solving the gap equation directly at complex momenta. A growing discrepancy between the two calculations provides a numerical diagnostic of the region in which the analytic continuation becomes unreliable and may indicate the proximity of nonanalytic structures.

To construct an analytic representation of the real-axis dressing functions, we use the network shown in Fig.~\ref{fig:diagram}C, whose GELU activation function is entire.
\begin{figure}[!bp]
\centering
\includegraphics[width=\columnwidth]{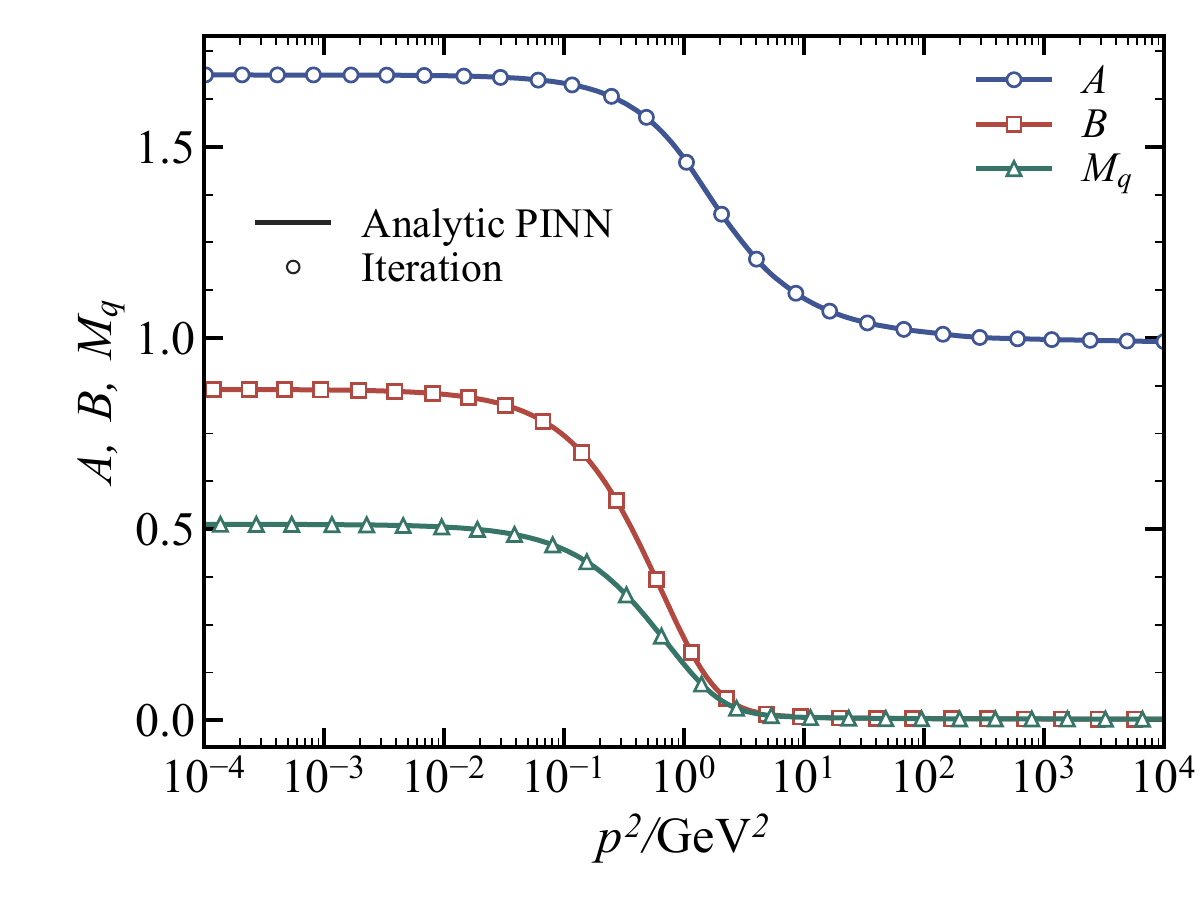}
\caption{Solutions of the quark gap equation on the real momentum axis. Open blue circles, red squares, and green triangles show the conventional iterative results for $A(p^2)$, $B(p^2)$, and $M_q(p^2)$, respectively. Solid curves of the same colors show the corresponding analytic PINN solutions. The function $A$ is dimensionless, and $B$ and $M_q$ are given in $\mathrm{GeV}$.
\label{fig4}}
\end{figure}

\begin{figure}[!t]
\centering
\begin{overpic}[width=0.92\linewidth]{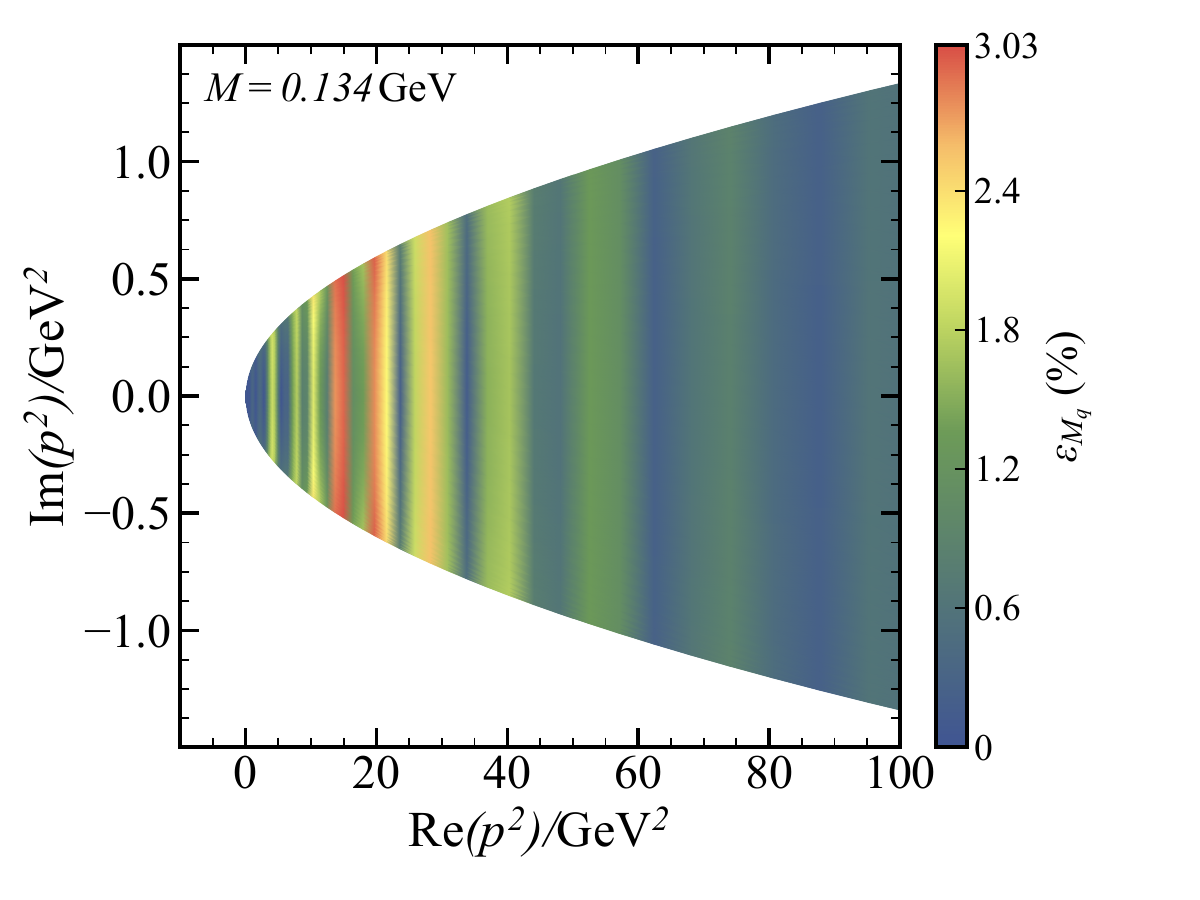}
\end{overpic}
\begin{overpic}[width=0.92\linewidth]{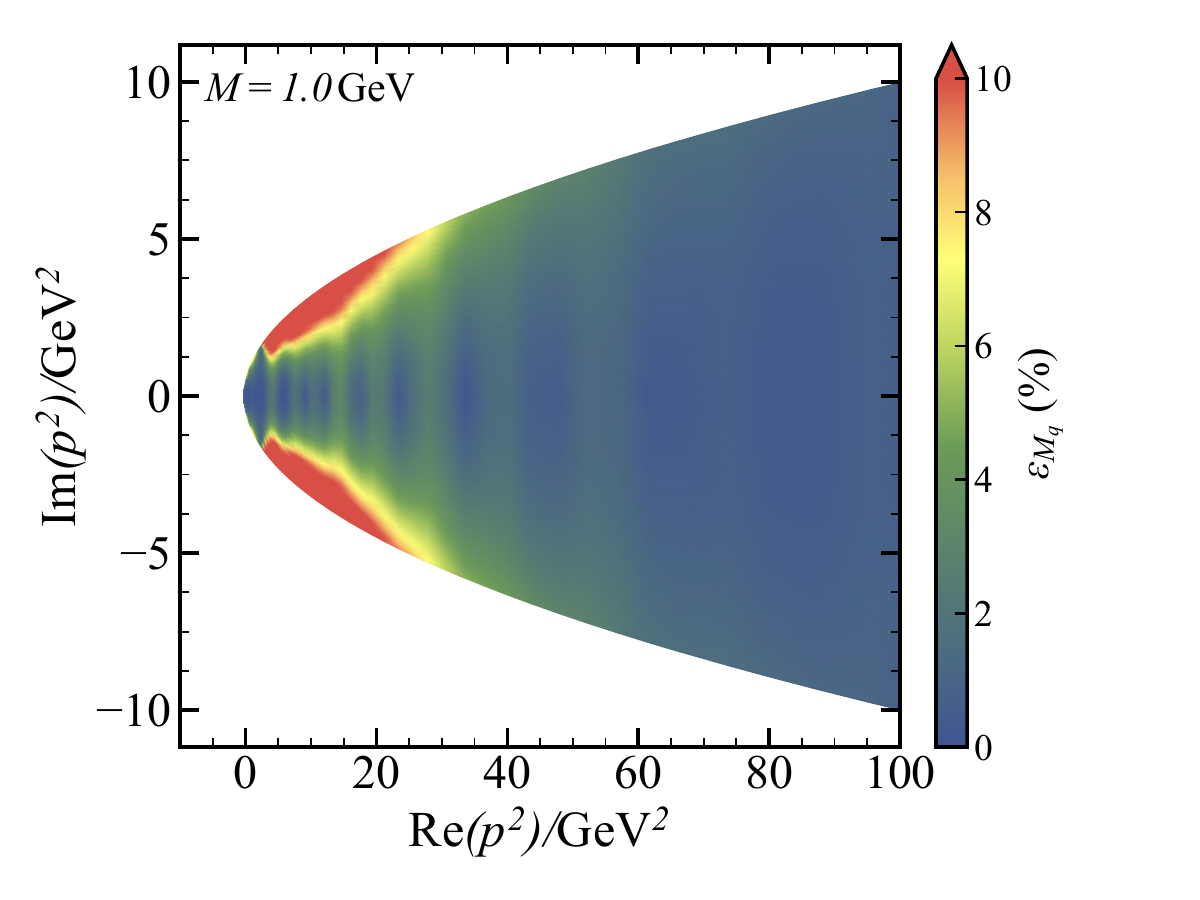}
\end{overpic}
\begin{overpic}[width=0.92\linewidth]{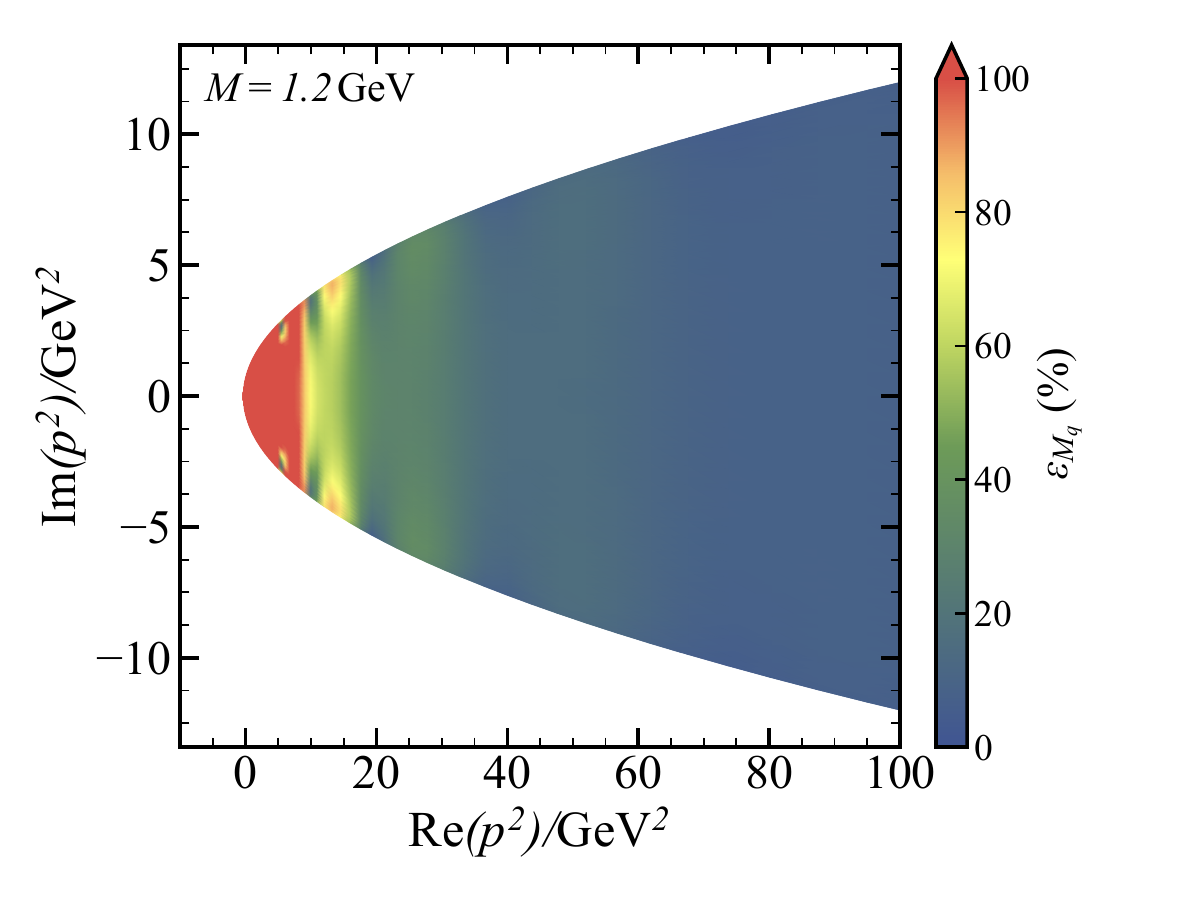}
\end{overpic}
\caption{Relative difference $\epsilon_{M_q}(p^2)$ for the quark mass function $M_q$. From top to bottom, $M=0.134$, $1.0$, and $1.2~\mathrm{GeV}$. Increasing $M$ enlarges the parabolic domain. The color scale progresses from blue through green and yellow to red as the relative difference increases. Triangular tips indicate values above the displayed range.
\label{fig5}}
\end{figure}
\begin{table*}[!t]
\caption{\label{table1}
Pole positions $q$ and corresponding residues $(R_V,R_S)$ for the three complex-conjugate pairs obtained by analytic continuation of the real-axis PINN solution.
For each pair, $M_{\mathrm{crit}}$ is the value of $M$ at which the
parabolic contour reaches the poles.}
\begin{ruledtabular}
\begin{tabular}{cccc}
$q$ [$\mathrm{GeV}^2$] & $M_{\mathrm{crit}}$ [$\mathrm{GeV}$]
& $R_V$ & $R_S$ [$\mathrm{GeV}$] \\
\hline
$-0.163\pm0.421i$           & $1.109$ & $\phantom{-}0.0835\mp0.4127i$ & $-0.1104\mp0.2605i$ \\
$\phantom{-}0.081\pm0.887i$ & $1.273$ & $-0.0109\mp0.0182i$           & $\phantom{-}0.0196\pm0.0040i$ \\
$\phantom{-}0.107\pm0.975i$ & $1.322$ & $-0.0310\mp0.0234i$           & $-0.0377\pm0.0074i$ \\
\end{tabular}
\end{ruledtabular}
\end{table*}
The dressing functions obtained on the real momentum axis are shown in Fig.~\ref{fig4}. As can be seen, the analytic PINN solutions remain in excellent agreement with those obtained using the conventional iterative method, with mean relative errors of $0.0151\%$ for $A(p^2)$ and $0.195\%$ for $B(p^2)$, and $0.192\%$ for $M_q(p^2)$. This demonstrates that the PINN framework retains a certain degree of flexibility with respect to the choice of network architecture and activation function while maintaining high numerical accuracy.

To continue the real-axis dressing functions analytically, we evaluate the trained PINN at complex values of $p^2$ without further training. We quantify the difference between the continued dressing functions and those computed directly at complex momenta using $\epsilon_{M_q}(p^2)=\left|{M_q}^{\mathrm{AC}}(p^2)-{M_q}^{\mathrm{ref}}(p^2)\right|/\left|{M_q}^{\mathrm{ref}}(p^2)\right|$, where  ${M_q}^{\mathrm{ref}}(p^2)$ denotes the corresponding solution in Section~\ref{subsec:SC}. The resulting relative differences are shown in Fig.~\ref{fig5}.

From top to bottom, the three sets of panels correspond to $M=0.134$, $1.0$ and $1.2~\mathrm{GeV}$, respectively. For $M=0.134~\mathrm{GeV}$, the two calculations agree closely over most of the parabolic domain. The mean relative difference between the analytically continued PINN solution and the direct DSE solution is $0.195\%$ for $M_q$ ($0.0155\%$ for $A$ and $0.198\%$ for $B$). These results show that the PINN solution on the real axis can be continued into the parabolic domain corresponding to the pion mass with percent-level or better accuracy.

As the parabolic domain is enlarged to $M=1.0~\mathrm{GeV}$, the discrepancies increase but remain at the level of a few percent over most of the domain. However, the growth of the discrepancy near the infrared boundary suggests that the expanding contour is approaching a pair of complex-conjugate singularities.

At $M=1.2~\mathrm{GeV}$, the direct DSE solution disagrees with the analytically continued PINN solution across nearly the entire infrared region. The relative difference in $M_q$ exceeds $10^3\%$ at every sampled point with $\operatorname{Re}(p^2)<1~\mathrm{GeV}^2$. This widespread loss of agreement suggests that a nonanalytic structure has entered the contour, invalidating the analyticity assumption required to reconstruct the dressing functions from their boundary values using Eq.~(\ref{cauchy}).

These deviations suggest that the expanding parabolic contour first encounters the leading complex-conjugate singularities between $M=1.0$ and $1.2~\mathrm{GeV}$.

The analytic PINN representation also allows us to locate the propagator poles directly. We continue the real-axis dressing functions $A$ and $B$ to complex $q=p^2$ and solve $D(q)=qA^2(q)+B^2(q)=0$. Table~\ref{table1} lists the three complex-conjugate pole pairs found in the sampled domain for $M<1.4~\mathrm{GeV}$, together with their vector and scalar residues. The pairs are ordered by the value of $M_{\mathrm{crit}}$ at which the expanding parabolic contour first reaches each pair. The first pair is located at $q=-0.163\pm0.421i~\mathrm{GeV}^2$ and enters the domain at $M_{\mathrm{crit}}\simeq1.109~\mathrm{GeV}$, within the range of $1.0$--$1.2~\mathrm{GeV}$ inferred from Fig.~\ref{fig5}.

We extracted the residues by integrating each propagator component around a small counterclockwise contour $\mathcal C_q$ enclosing only the pole at $q$, so that $R_X=(2\pi i)^{-1}\oint_{\mathcal C_q}\sigma_X(z)dz$ for $X\in\{V,S\}$. The residue ratio satisfies $(R_S/R_V)^2=[B(q)/A(q)]^2=-q$ for all three pole pairs. Moreover, for both the second and third pole pairs, the magnitudes of the vector and scalar residues are substantially smaller than those of the corresponding residues of the first pair. This residue hierarchy, together with the observation that the sampled infrared momenta lie closer to the first pole pair than to the other two, supports treating the first pair as the dominant pole contribution to the propagator in the sampled domain.

\subsection{Spectral reconstruction with a single complex-conjugate pole pair}\label{subsec:SRPP}

The pole positions and residues extracted by analytic continuation characterize the local behavior of the quark propagator near each pole. To determine how the pole and spectral contributions together reproduce its momentum dependence across the sampled domain, we employ a spectral representation with PINN-parametrized vector and scalar spectral densities.

It is convenient to express the quark propagator in terms of its vector and scalar components, $\sigma_V$ and $\sigma_S$, respectively:
\begin{align}
    S(p)=-i\gamma\cdot p\sigma_V(p^2)+\sigma_S(p^2),
\end{align}
where 
\begin{subequations}\label{relations}
\begin{align}
    \sigma_V(p^2)&=\frac{A(p^2)}{p^2A^2(p^2)+B^2(p^2)}\\
    \sigma_S(p^2)&=\frac{B(p^2)}{p^2A^2(p^2)+B^2(p^2)}.
\end{align}
\end{subequations}

\begin{figure*}[!t]
\centering
\begin{minipage}{0.49\textwidth}
    \centering
    \begin{overpic}[width=\linewidth]{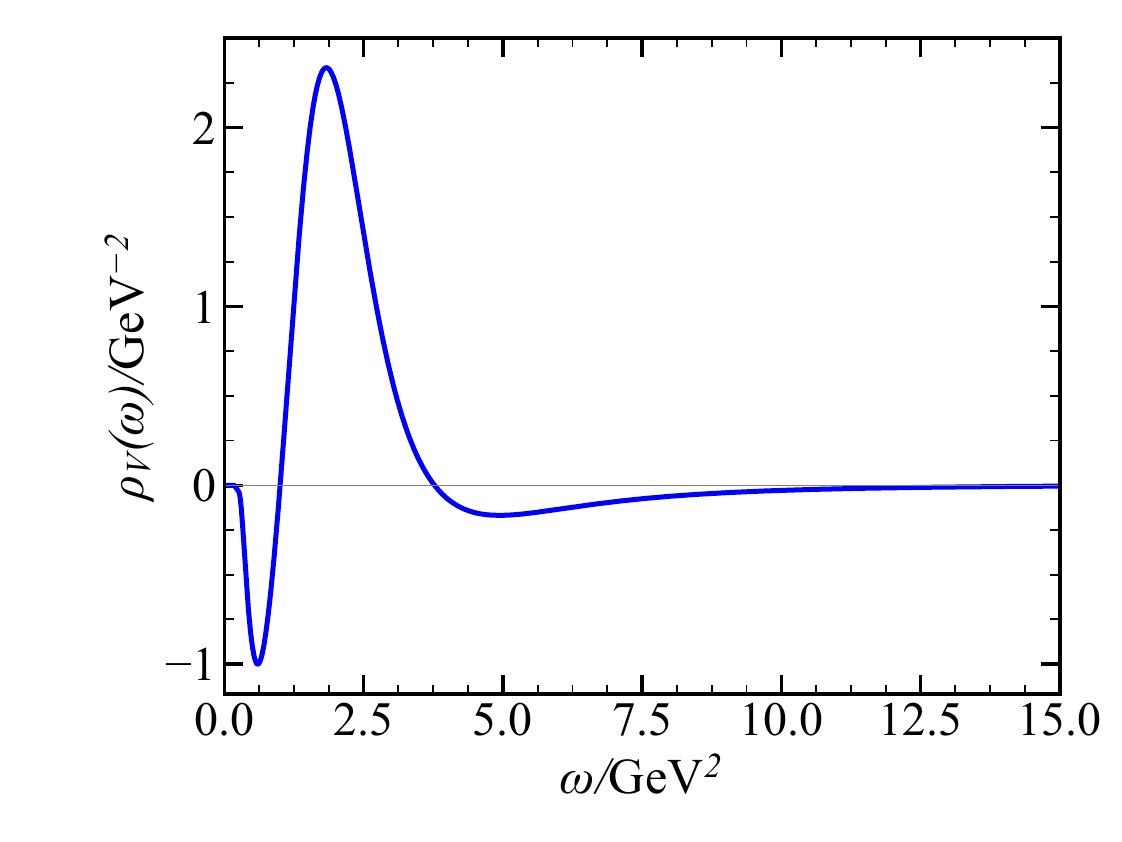}
    \put(2,65){\textsf{\large A}}
    \end{overpic}
    \begin{overpic}[width=\linewidth]{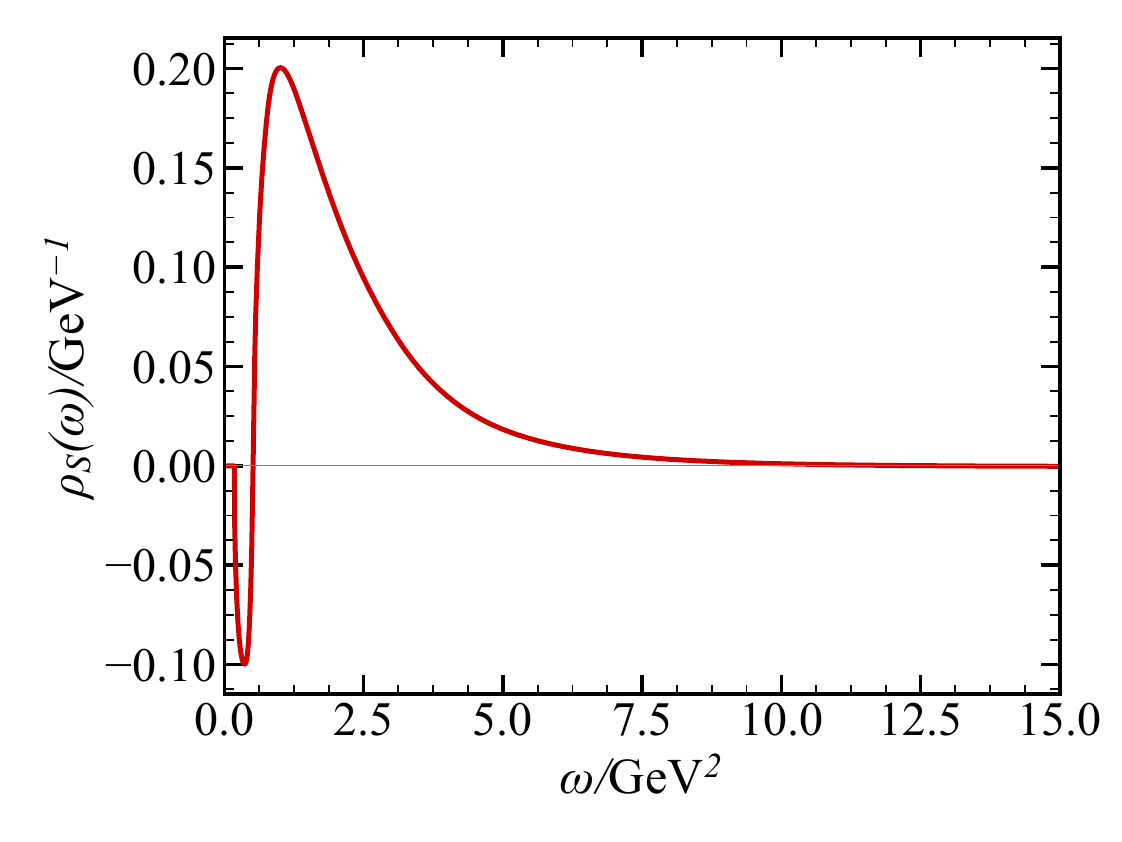}
    \put(2,65){\textsf{\large B}}
    \end{overpic}
\end{minipage}
\hfill
\begin{minipage}{0.49\textwidth}
    \centering
    \begin{overpic}[width=\linewidth]{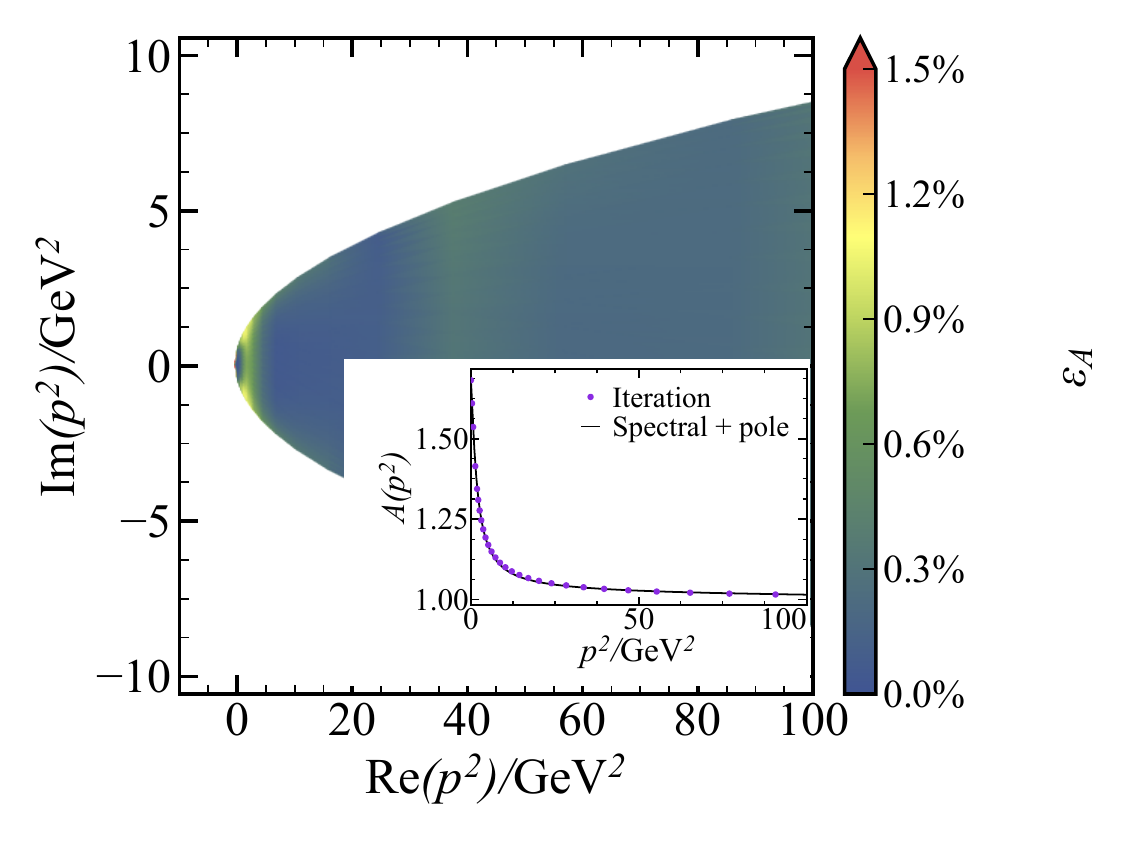}
    \put(2,65){\textsf{\large C}}
    \end{overpic}
    \begin{overpic}[width=\linewidth]{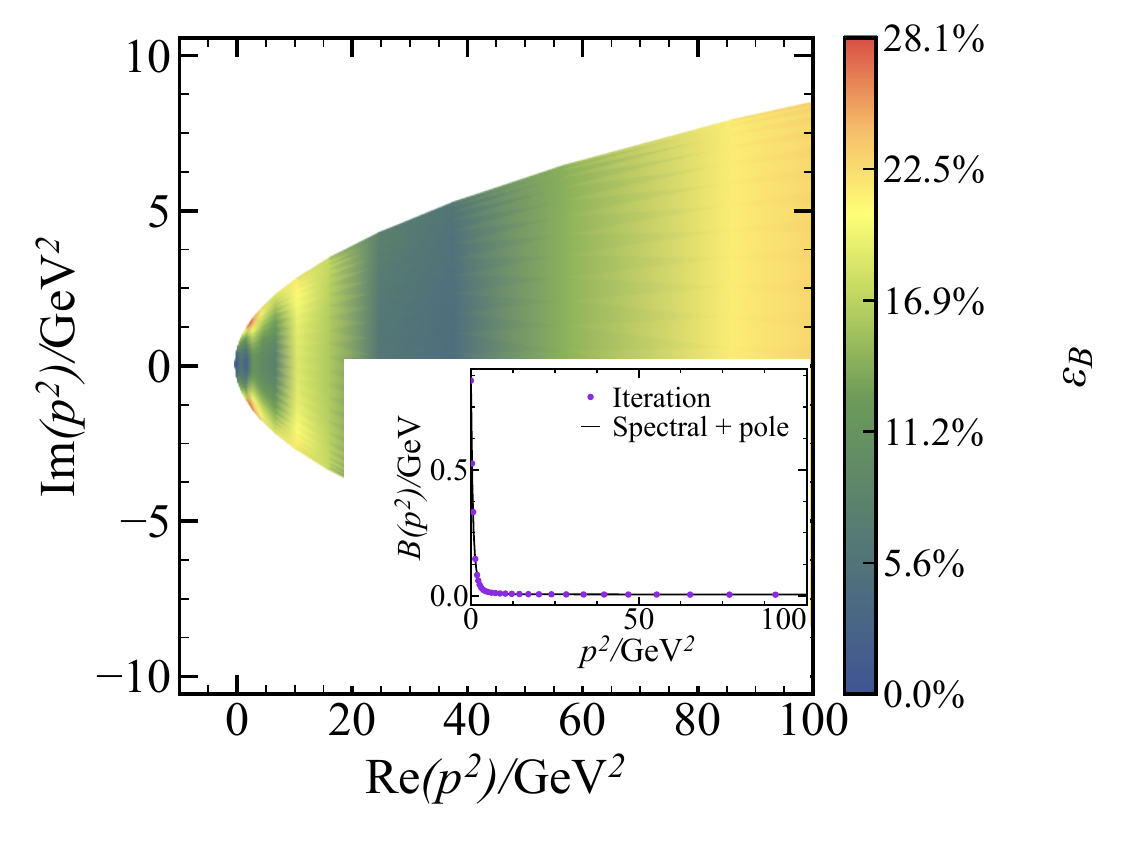}
    \put(2,65){\textsf{\large D}}
    \end{overpic}
\end{minipage}
\caption{Spectral reconstruction of the quark propagator. \textsf{Panels A, B.} show the vector and scalar spectral densities, $\rho_V(\omega)$ and $\rho_S(\omega)$, respectively. \textsf{Panels C, D.} show the relative errors $\epsilon_A(p^2)$ and $\epsilon_B(p^2)$ between the reconstructed and reference dressing functions. The insets show the corresponding comparisons on the real momentum axis.
\label{fig6}}
\end{figure*}

Using contour integration, we decompose the quark propagator into a contribution associated with the branch cut along the negative real axis and contributions from complex-conjugate poles in the complex momentum plane.
\begin{subequations}\label{sigma}
\begin{align}
    \sigma(p^2)&=\frac{1}{\pi}\int_0^\infty\frac{\rho(\omega)}{p^2+\omega}d\omega+\sum_i(\frac{R_i}{p^2-q_i}+\frac{R_i^*}{p^2-q_i^*}),\\
    \rho(\omega)&=\operatorname{Im}(\sigma(-\omega-i\epsilon)),
\end{align}
\end{subequations}

Here, $q_i$ and $R_i$ denote the pole locations and the corresponding residues, respectively; $\epsilon$ is a positive infinitesimal; and $\rho(\omega)$ is the spectral density. Previous studies have investigated the singularity structure of the quark propagator and identified multiple pairs of complex-conjugate singularities in the complex momentum plane~\cite{Windisch:2016iud,Kaidisch:2023lwp,Pawlowski:2024kxc}. Within the complex-momentum domain relevant to meson bound-state calculations, the dominant isolated contribution can be approximated by the leading complex-conjugate pole pair~\cite{Dorkin:2013rsa}. For residues of comparable magnitude, the conjugate pair with a pole closest to the evaluation point $p^2$ is expected to give the largest contribution to the pole sum. We therefore adopt a minimal representation consisting of one complex-conjugate pole pair and a continuous spectral contribution along the negative real axis. Eq.~(\ref{sigma}a) then becomes
\begin{align}\label{exchange}
    \sigma(p^2)=\frac{1}{\pi}\int_0^\infty\frac{\rho(\omega)}{p^2+\omega}d\omega+\frac{R}{p^2-q}+\frac{R^*}{p^2-q^*}.
\end{align}
Since the transformation from $\sigma_{S,V}$ to the dressing functions $A$ and $B$ can significantly amplify small numerical errors in $\sigma_{S,V}$, an accurate reconstruction of the propagator requires particularly high precision. An accurate fit over a finite momentum range does not guarantee the correct ultraviolet behavior. We therefore explicitly incorporate the known ultraviolet asymptotic behavior into the network ansatz, thereby constraining the functional form represented by the neural network.

For a nonzero current-quark mass, the quark dressing functions have the following asymptotic forms as $p^2\to\infty$~\cite{Roberts:1994dr}.
\begin{subequations}
\begin{align}
        A(p^2)&\to 1,\\
        B(p^2)&\to \frac{\hat{m}}{L_p(p^2)^{\gamma_m}},
\end{align}
\end{subequations}
where $L_p(p^2)=\frac{1}{2}\ln\left(\frac{p^2}{\Lambda_{QCD}^2}\right)$ and $\hat{m}$ denotes the renormalization-group-invariant current-quark mass. Substituting these ultraviolet asymptotic forms into Eq.~(\ref{relations}), we obtain the corresponding ultraviolet behavior of the $\sigma_{S,V}$ functions:
\begin{subequations}\label{sigmavs}
    \begin{align}
        \sigma_V(p^2)&\to\frac{1}{p^2},\\
        \sigma_S(p^2)&\to\frac{\hat{m}}{p^2L_p(p^2)^{\gamma_m}}.
    \end{align}
\end{subequations}
Combining these limits with Eq.~(\ref{exchange}) gives two sum rules relating the spectral densities to the pole residues.
\begin{subequations}\label{rer}
    \begin{align}
        \frac{1}{\pi}\int_0^{+\infty}\rho_V(\omega)d\omega + 2 \operatorname{Re}(R_V)&=\lim_{p^2\to\infty}p^2\sigma_V(p^2)=1,\\
        \frac{1}{\pi}\int_0^{+\infty}\rho_S(\omega)d\omega + 2 \operatorname{Re}(R_S)=&\lim_{p^2\to\infty}p^2\sigma_S(p^2)=0.
    \end{align}
\end{subequations}
Equation~(\ref{sigma}b) also determines the large-$\omega$ behavior of the vector and scalar spectral densities.
\begin{subequations}
    \begin{align}
    \rho_V(\omega)&\to 0,\\
        \rho_S(\omega)&\to -\frac{\hat{m}\gamma_m\pi}{2\omega L_\omega(\omega)^{\gamma_m+1}},
    \end{align}
\end{subequations}
where $L_\omega(\omega)=\frac{1}{2}\ln\left(\frac{\omega}{\Lambda_{QCD}^2}\right)$.

Motivated by the analyticity required by Cauchy's integral formula~\cite{Serna:2020txe}, we assume that the branch cut does not enter the interior of the parabolic computational domain. We therefore introduce a finite spectral threshold
$\omega_{\mathrm{th}}=\eta^2M^2$ and impose
\begin{align}
\rho_{V,S}(\omega)=0,
\qquad
0\leq\omega\leq\omega_{\mathrm{th}}.
\end{align}

Combining the above constraints, we adopt the following ansatz for the spectral density $\rho(\omega)$,
\begin{subequations}
    \begin{align}
        \rho_V(\omega)&=f_{\mathrm{ir}}(\omega)f_{\mathrm{uv}}^V(\omega)N_V(\omega),\\
        \rho_S(\omega)&=f_{\mathrm{ir}}(\omega)f_{\mathrm{uv}}^S(\omega)\left[1+f_{\mathrm{match}}(\omega)N_S(\omega)\right],
    \end{align}
\end{subequations}
where $N_V(\omega)$ and $N_S(\omega)$ are neural networks that describe the remaining freedom in the vector and scalar spectral densities, respectively. The prefactors enforce the threshold and ultraviolet constraints. Their explicit forms and parameter values are given in Appendix~\ref{Output layer}. We set the contour parameter to $M=0.85~\mathrm{GeV}$ so that the leading complex-conjugate poles lie far away from the parabolic integration domain, avoiding numerical contamination of the reference solution at complex momenta from a pole approaching or entering the contour.

\begin{table}[t]
\caption{\label{table2}
Location and residues of the fitted complex-conjugate pole
pair.}
\begin{ruledtabular}
\begin{tabular}{ccc}
$q~[\mathrm{GeV}^2]$ &
$R_V$ &
$R_S~[\mathrm{GeV}]$\\
\hline
$-0.146\pm0.391i$ &
$0.147\mp0.400i$ &
$-0.062\mp0.269i$
\end{tabular}
\end{ruledtabular}
\end{table}

We employ the neural-network architecture illustrated in Fig.~\ref{fig:diagram}D and treat $\operatorname{Re}(q)$, $\operatorname{Im}(q)$, $\operatorname{Im}(R_V)$, and $\operatorname{Im}(R_S)$ as additional trainable parameters, which are optimized simultaneously with the neural-network parameters. The spectral sum rules in Eq.~(\ref{rer}) determine $\operatorname{Re}(R_V)$ and $\operatorname{Re}(R_S)$. The reconstructed spectral densities are presented in Fig.~\ref{fig6}, while the extracted pole locations and residues are listed in Table~\ref{table2}. Together with the fitted pole pair, the reconstructed spectral densities reproduce the momentum dependence of both $A(p^2)$ and $B(p^2)$ on the real axis, as shown by the comparison with the numerical solutions of the gap equation in the insets of Fig.~\ref{fig6}C and D. The spectral representation provides a good overall fit to both dressing functions, with mean relative errors of $0.282\%$ for $A(p^2)$ and $1.40\%$ for $B(p^2)$. However, the larger deviations observed in $B(p^2)$ in some momentum regions may partly reflect the restriction of the spectral representation to a single complex-conjugate pole pair.

The vector and scalar spectral densities reported in Ref.~\cite{Zhu:2020bwu} exhibit relatively simple peak--minimum structures, whereas our reconstruction yields several alternating positive and negative regions in both densities. This sign-indefinite behavior, particularly the regions where $\rho_V$ is negative, is incompatible with a conventional positive K\"{a}ll\'{e}n--Lehmann representation and signals violation of reflection positivity. Together with the complex-conjugate pole pair, it indicates that the dressed-quark propagator cannot be associated with a stable asymptotic particle state, consistent with quark confinement. The pronounced peaks occur at intermediate timelike scales, around $\omega\simeq1.83~\mathrm{GeV}^2$ for $\rho_V$ and $\omega\simeq1.01~\mathrm{GeV}^2$ for $\rho_S$, corresponding to $\sqrt{\omega}=\mathcal{O}(1~\mathrm{GeV})$. These peaks are not physical mass shells. They are concentrated continuum contributions which, after cancellation against the neighboring negative regions and the pole terms, reproduce the Euclidean propagator.

On the spacelike axis, the two pole terms are complex conjugates, so their sum is real for each propagator component. A check on the fitted residues is provided by the relation $(R_S/R_V)^2=-q$. The spectral fit leaves this relation unconstrained. Nevertheless, the relative difference of these two values is $5.5\%$, indicating that the fitted vector and scalar residues approximately satisfy the expected relation.

Table~\ref{table1} allows a separate comparison with the first pole pair obtained by direct analytic continuation. Relative to the values in that table, the fitted magnitudes $|R_V|$ and $|R_S|$ differ by only $1.1\%$ and $2.5\%$, respectively. The corresponding phase differences are $8.7^\circ$ for $R_V$ and $9.9^\circ$ for $R_S$. The pole positions obtained by spectral fitting and by direct analytic continuation correspond to $M_{\mathrm{crit}}=1.062$ and $1.109~\mathrm{GeV}$, respectively. Relative to the spectral-fit values, the pole position obtained by direct analytic continuation differs by $8.3\%$ and the corresponding contour threshold by $4.4\%$. The agreement in pole position and residue magnitudes, together with the reconstruction accuracy shown in Fig.~\ref{fig6}, supports treating the first conjugate pair as the dominant isolated-pole contribution in the sampled momentum domain.

\subsection{Pole and spectral contributions to the order parameters}\label{subsec:OP}

The chiral condensate $\langle\bar{\psi}\psi\rangle$ and the quark number density $\langle \psi^\dagger \psi\rangle$ have been used as complementary order parameters in effective-potential studies of the QCD phase structure~\cite{Li:2025cll,Brodsky:2012ku,Gao:2015kea,klevansky1992nambu,Farias:2006cs,Chen:2008zr}. Here we use the spectral representation of the vacuum $N^+$ quark propagator to determine how its analytic structure contributes to these quantities at zero temperature and finite quark chemical potential $\mu>0$. Following Ref.~\cite{Zong:2008sm}, we shift the Euclidean energy according to $p_4\to p_4+i\mu$ and evaluate the change in each quantity relative to its vacuum value. For a single quark flavor,
\begin{subequations}
\begin{align}
    \langle\bar{\psi}\psi\rangle_{\mu}
    &=-N_c Z_4\int_p
      \mathrm{tr}_D\!\left[S(p;\mu)\right],\\
    \langle\psi^\dagger\psi\rangle_{\mu}
    =\langle\bar{\psi}\gamma_4\psi\rangle_{\mu}
    &=-N_c Z_2\int_p
      \mathrm{tr}_D\!\left[\gamma_4 S(p;\mu)\right],
\end{align}
\end{subequations}
where $\int_p\equiv\int d^4p/(2\pi)^4$ and $\mathrm{tr}_D$ denotes the Dirac trace.

We define the medium contributions by subtracting the corresponding vacuum values. For the shifted vacuum $N^+$ propagator, this subtraction gives
\begin{subequations}
\begin{align}
    &\Delta\langle\bar{\psi}\psi\rangle(\mu)
    \equiv \langle\bar{\psi}\psi\rangle_{\mu}
          - \langle\bar{\psi}\psi\rangle_{0}
          \notag\\
    &=-4N_c Z_4 \int_{\vec q}
      \left(\int_{C_{\mu}}-\int_{C_0}\right)
      \frac{dz}{2\pi}\,\sigma_S(k^2+z^2),\\
    &\Delta\langle\psi^\dagger\psi\rangle(\mu)
    \equiv \langle\psi^\dagger\psi\rangle_{\mu}
          - \langle\psi^\dagger\psi\rangle_0
          \notag\\
    &=4iN_c Z_2 \int_{\vec q}
      \left(\int_{C_{\mu}}-\int_{C_0}\right)
      \frac{dz}{2\pi}\,z\,\sigma_V(k^2+z^2).
\end{align}
\end{subequations}
Here $k=|\vec q|$, $\int_{\vec q}\equiv\int d^3q/(2\pi)^3$, and $C_a$ runs from $-\infty+ia$ to $+\infty+ia$. The shift $z=q_4+i\mu$ thus moves the energy integration from the real axis $C_0$ to the parallel contour $C_{\mu}$.
Substituting the spectral representations of $\sigma_S$ and $\sigma_V$ from Eq.~(\ref{exchange}) into the contour integrals, we evaluate separately the spectral and complex-conjugate-pole contributions to $\Delta\langle\bar{\psi}\psi\rangle(\mu)$ and $\Delta\langle\psi^\dagger\psi\rangle(\mu)$.
The spectral contributions to the chiral condensate and quark number density are
\begin{subequations}
\begin{align}
&\Delta\langle\bar{\psi}\psi\rangle_{\mathrm{spectral}}(\mu)
    =\frac{N_c Z_4}{2\pi^3}
      \int_0^{\mu^2} d\omega\,\rho_S(\omega)
      \notag\\
    &\qquad\times
      \left[
      \mu\sqrt{\mu^2-\omega}
      -\omega\ln\!\left(
      \frac{\mu+\sqrt{\mu^2-\omega}}{\sqrt{\omega}}
      \right)
      \right],\\
    &\Delta\langle\psi^\dagger\psi\rangle_{\mathrm{spectral}}(\mu)
    =\frac{N_c Z_2}{3\pi^3}
      \int_0^{\mu^2} d\omega\,\rho_V(\omega)
      (\mu^2-\omega)^{3/2}.
\end{align}
\end{subequations}

The complex-conjugate pole pair contributes
\begin{subequations}
\begin{align}
    &\Delta\langle\bar{\psi}\psi\rangle_{\mathrm{pole}}(\mu)
    =\frac{N_c Z_4}{\pi^2}\operatorname{Re}\!\left\{
      R_S\!\left[
      k_q E_q+q\ln\!\left(\frac{k_q+E_q}{\sqrt{-q}}\right)
      \right]\right\},\\
    &\Delta\langle\psi^\dagger\psi\rangle_{\mathrm{pole}}(\mu)
    =\frac{2N_c Z_2}{3\pi^2}\operatorname{Re}(R_V)\,k_q^3 ,
\end{align}
\end{subequations}
where $E_q=\sqrt{k_q^2-q}$ and $k_q=\left[\mu^2+\operatorname{Re}q-\frac{(\operatorname{Im}q)^2}{4\mu^2}\right]_+^{1/2},\qquad [x]_+\equiv\max(x,0)$. The pole pair begins to contribute when $k_q>0$, which requires
\begin{align}
    \mu>\mu_{\mathrm{pole}}
    \equiv\sqrt{\frac{|q|-\operatorname{Re}q}{2}}.
\end{align}

\begin{figure}[t]
\centering
\begin{overpic}[width=\columnwidth]{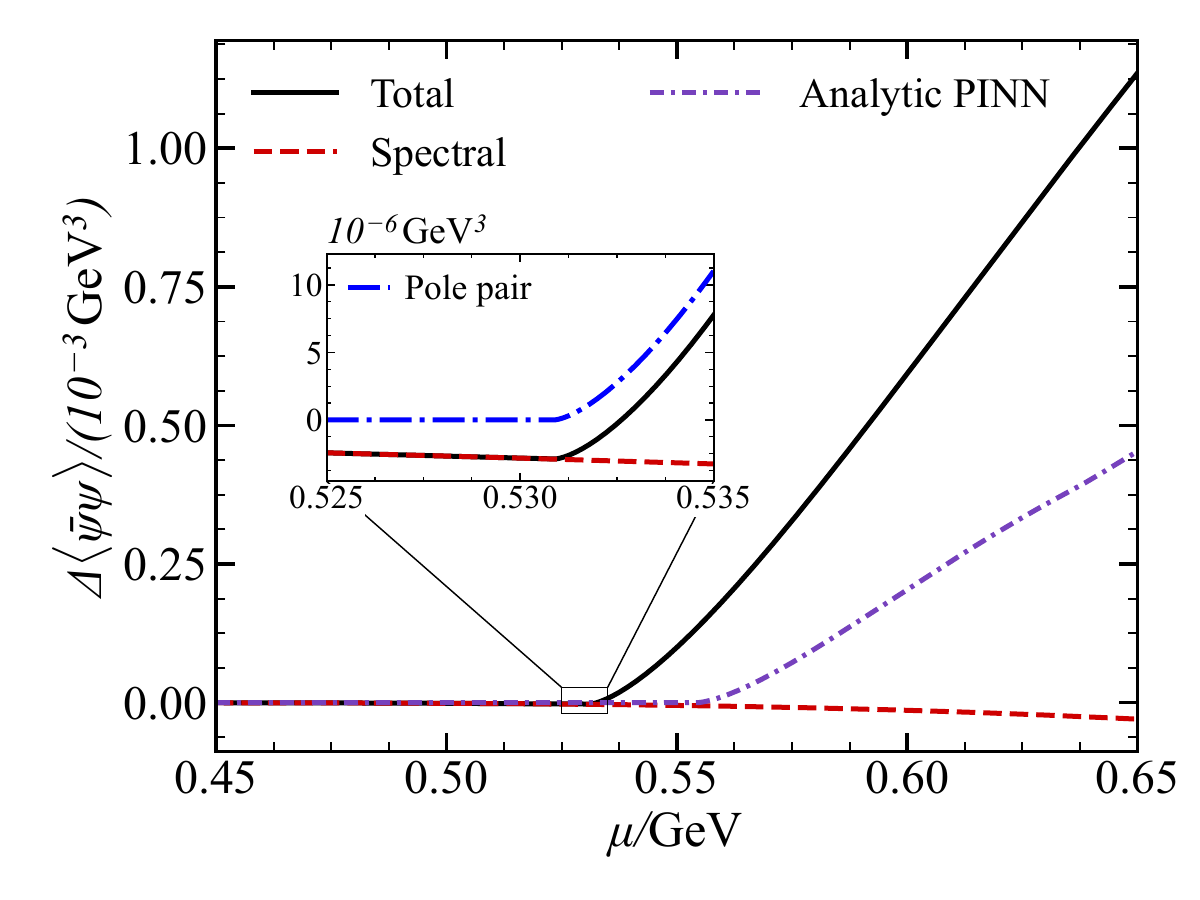}
    \put(2,68){\textsf{\large A}}
\end{overpic}
\\
\begin{overpic}[width=\columnwidth]{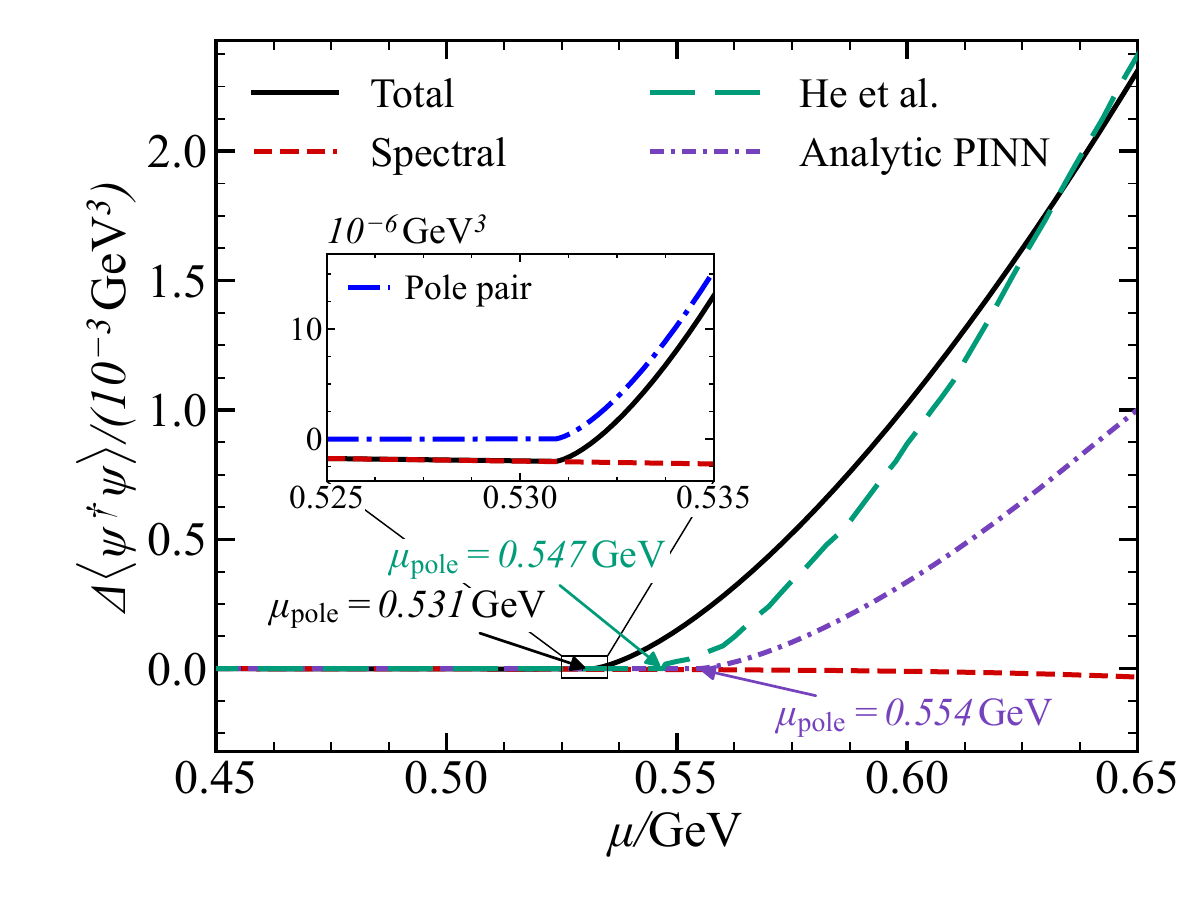}
    \put(2,68){\textsf{\large B}}
\end{overpic}
\caption{Vacuum-subtracted chiral condensate (\textsf{Panel A}) and quark number density (\textsf{Panel B}) for a single quark flavor on the $N^+$ branch at zero temperature and finite quark chemical potential $\mu$. For the spectral reconstruction, the solid black curves show the sum of the pole and spectral contributions, the dashed red curves show the spectral contribution, and the blue dash-dotted curves in the insets show the contribution from the fitted pole pair. The insets resolve the small negative spectral contribution near the fitted pole threshold. The purple short dash-dotted curves show the summed contributions from the three conjugate pole pairs obtained by direct analytic continuation of the real-axis PINN solution. In panel B, the green long-dashed curve shows the density digitized from Fig.~1 of Ref.~\cite{He:2009zzp} and divided by two to obtain the density per flavor.
\label{fig7}}
\end{figure}

Figure~\ref{fig7} shows the vacuum-subtracted chiral condensate (A) and quark number density (B) as functions of the quark chemical potential $\mu$ at zero temperature. Using the expressions above, we evaluate the spectral and pole-pair contributions from the spectral reconstruction of the vacuum $N^+$ propagator separately. We compare their sum with the combined contribution of the three pole pairs extracted by direct analytic continuation of the real-axis PINN solution.

In the spectral reconstruction framework, our pole threshold, $\mu_{\mathrm{pole}}\simeq0.531~\mathrm{GeV}$, falls within the range $0.512$--$0.630~\mathrm{GeV}$ obtained by Bai and Liu~\cite{Bai:2023ctq}, who numerically located the first zero of the quark propagator denominator for three unmodified DSE truncations. Below the pole threshold, only the continuous spectrum can contribute. Once $\mu^2$ exceeds $\omega_{\mathrm{th}}$, the integrals over the spectral densities sample the interval $\omega_{\mathrm{th}}<\omega<\mu^2$, where both $\rho_V$ and $\rho_S$ are negative and the integration weights are positive. This gives small negative changes in both the quark number density and the chiral condensate. Above $\mu_{\mathrm{pole}}$, the complex-conjugate pole pair gives a positive contribution that rapidly dominates the medium-induced changes in both the quark number density and the chiral condensate. At $\mu=0.60~\mathrm{GeV}$, the spectral contributions amount to only $-0.98\%$ of the total change in the quark number density and $-2.30\%$ of the total change in the chiral condensate. The pole pair alone therefore reproduces both medium-induced changes on the vacuum $N^+$ branch to within a few percent at this chemical potential. At $\mu=0.65~\mathrm{GeV}$, our single-flavor quark number density agrees to within $2\%$ with the value obtained from Eq.~(16) of He et al.~\cite{He:2009zzp}, who represented the vacuum quark propagator by three pairs of complex-conjugate poles.

The purple curves obtained by direct analytic continuation rise more slowly than the black curves from the spectral reconstruction. The first pole pair enters the integration domain at a higher chemical potential $\mu_{\mathrm{pole}}=0.554~\mathrm{GeV}$, leaving a smaller momentum range $0<k<k_q$ at the same $\mu$. At $\mu=0.60~\mathrm{GeV}$, the slopes of both purple curves are approximately half those of the corresponding black curves.

Representative zero-temperature DSE calculations place the chiral transition at quark chemical potentials of about $0.3~\mathrm{GeV}$~\cite{Qin:2010nq}, below the pole threshold $\mu_{\mathrm{pole}}$ found here. Our finite-density analysis follows only the vacuum $N^+$ branch, and $\mu_{\mathrm{pole}}$ marks the point at which the shifted energy contour first reaches its complex-conjugate pole pair. Determining the chiral-transition chemical potential for our parameter set would require comparing the thermodynamic potentials of the competing phases, in particular the Nambu and Wigner phases, to identify the equilibrium phase with the lowest thermodynamic potential. Below $\mu_{\mathrm{pole}}$, both the quark number density and the chiral condensate remain close to their vacuum values, indicating that the Silver Blaze property is approximately satisfied on the $N^+$ branch over this range.

\section{Summary and perspective}\label{sec:summary}

In this work, we developed a physics-informed neural-network framework for solving the quark Dyson--Schwinger equation both on the real momentum axis and in the complex momentum plane. By embedding the gap equation directly into the loss function, we can represent the dressing functions as continuous and differentiable functions. The agreement with results from the conventional iterative method demonstrates that this approach provides a reliable alternative representation of the quark propagator. Moreover, with suitable initialization, the framework resolves the coexisting Nambu and Wigner branches, making it useful for nonlinear DSE systems with multiple solutions.

We trained an analytic neural network to represent the dressing functions by imposing the quark gap equation on the real momentum axis. Within small parabolic domains, the continued dressing functions agree closely with those calculated directly at complex momenta. The growing discrepancy as the domain expands suggests that the parabolic boundary reaches the leading pole pair at a value of $M$ between $1.0$ and $1.2~\mathrm{GeV}$. We cross-checked this estimate by solving $qA^2(q)+B^2(q)=0$ with the analytically continued dressing functions and found three complex-conjugate pole pairs with $M_{\mathrm{crit}}<1.4~\mathrm{GeV}$ in the sampled domain. The expanding parabolic contour first reaches the pair at $q=-0.163\pm0.421i~\mathrm{GeV}^2$ when $M_{\mathrm{crit}}\simeq1.109~\mathrm{GeV}$, within the estimated range of $1.0$--$1.2~\mathrm{GeV}$. The residues extracted at all three pairs satisfy $(R_S/R_V)^2=-q$ to numerical precision. For the second and third pairs, the vector and scalar residue magnitudes are at most $9.2\%$ and $13.6\%$, respectively, of those of the first pair. This hierarchy, together with the proximity of the first pair to the sampled infrared momenta, supports its dominance among the isolated-pole contributions in that region.

The spectral reconstruction of the vacuum $N^+$ propagator yields sign-changing vector and scalar spectral densities and a dominant complex-conjugate pole pair at $q=-0.146\pm0.391i~\mathrm{GeV}^2$. The negative regions of $\rho_V$ and the complex poles indicate a violation of reflection positivity, consistent with the confinement interpretation of the dressed-quark propagator. The expanding parabolic domain first encounters the fitted pole pair at $M_{\mathrm{crit}}\simeq1.062~\mathrm{GeV}$, close to the threshold of $1.109~\mathrm{GeV}$ obtained by direct analytic continuation. The two pole positions differ by only $8.3\%$ of the fitted pole modulus. The fitted magnitudes of $R_V$ and $R_S$ differ from the direct-continuation values by only $1.1\%$ and $2.5\%$, respectively, while their phases differ by $8.7^\circ$ and $9.9^\circ$. This agreement in pole position and residue magnitudes, together with the successful reconstruction of both dressing functions, supports treating the leading complex-conjugate pair as the dominant isolated-pole contribution in the sampled momentum domain.

At zero temperature and finite quark chemical potential, we use the reconstructed vacuum $N^+$ propagator to separate the pole and spectral contributions to the quark number density and chiral condensate. Below $\mu_{\mathrm{pole}}\simeq0.531~\mathrm{GeV}$, both quantities remain close to their vacuum values, consistent with approximate Silver Blaze behavior on this branch. Once the shifted energy contour crosses the fitted pole pair, its contribution rapidly dominates both medium-induced changes. In a complementary calculation using the three pole pairs obtained by direct analytic continuation, the first pair begins to contribute at $\mu\simeq0.554~\mathrm{GeV}$. This later onset reduces the available momentum phase space at a given $\mu$, contributing to the slower rise of both observables. Locating the chiral transition requires comparing the thermodynamic potentials of the $N^+$, $N^-$, and Wigner branches to identify the stable phase at each chemical potential.

Future work should assess the robustness of the spectral reconstruction against more general singularity Ans\"atze, different interaction kernels, and truncations beyond rainbow--ladder. The resulting complex-momentum propagator can then be applied directly to meson Bethe--Salpeter equations to investigate the influence of explicit pole contributions on heavier bound states. By tracking the quark propagator singularities and comparing the thermodynamic potentials of the Nambu and Wigner branches in the $(T,\mu)$ plane, we could further investigate the nature of the QCD chiral transition.
\appendix

\section{\label{Appendix A}The neural-network architectures}

The core of a PINN is a feedforward neural network that maps the input variables to the desired output functions. For the Dyson--Schwinger equations considered here, the input contains the relevant momentum variables, while the outputs are the propagator dressing functions. The network architectures employed in this work are summarized in Fig.~\ref{fig:diagram}.

\subsection{\label{Input layer}Input layer}

The input layer receives and preprocesses the physical variables supplied to the neural network. For the solution on the real momentum axis, the network takes the squared Euclidean momentum $p^2$ as its only input. For the solution in the complex momentum plane, the squared momentum $p_+^2=p^2-\eta^2M^2+2i\eta M \sqrt{p^2}z_p$ is parametrized by two real inputs, $p^2$ and the angular variable $z_p$ in the meson rest frame.

\begin{figure*}[t]
\centering
\begin{overpic}[width=\columnwidth]{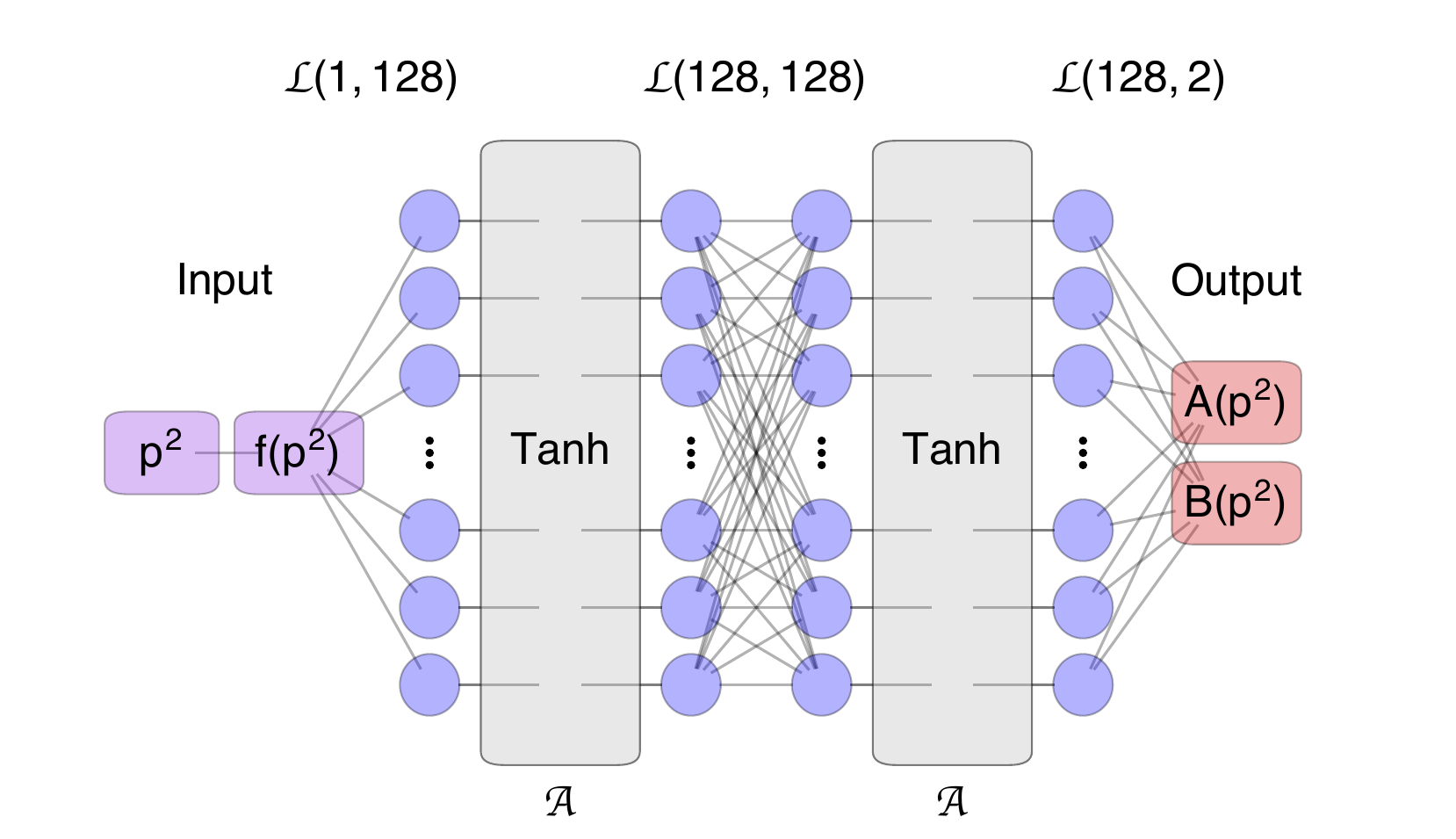}
    \put(2,55){\textsf{\large A}}
\end{overpic}
\begin{overpic}[width=\columnwidth]{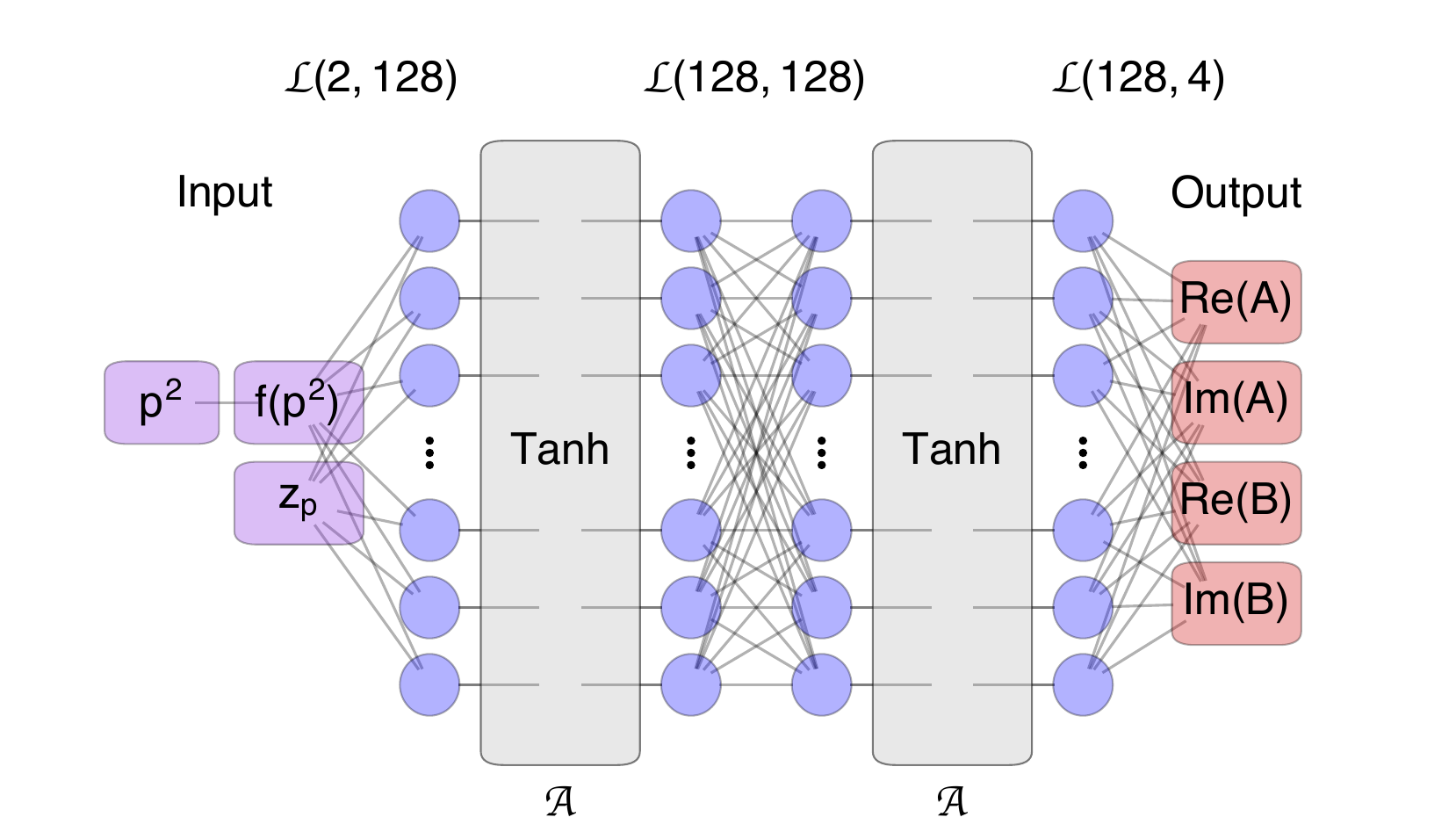}
    \put(2,55){\textsf{\large B}}
\end{overpic}
\\
\begin{overpic}[width=\columnwidth]{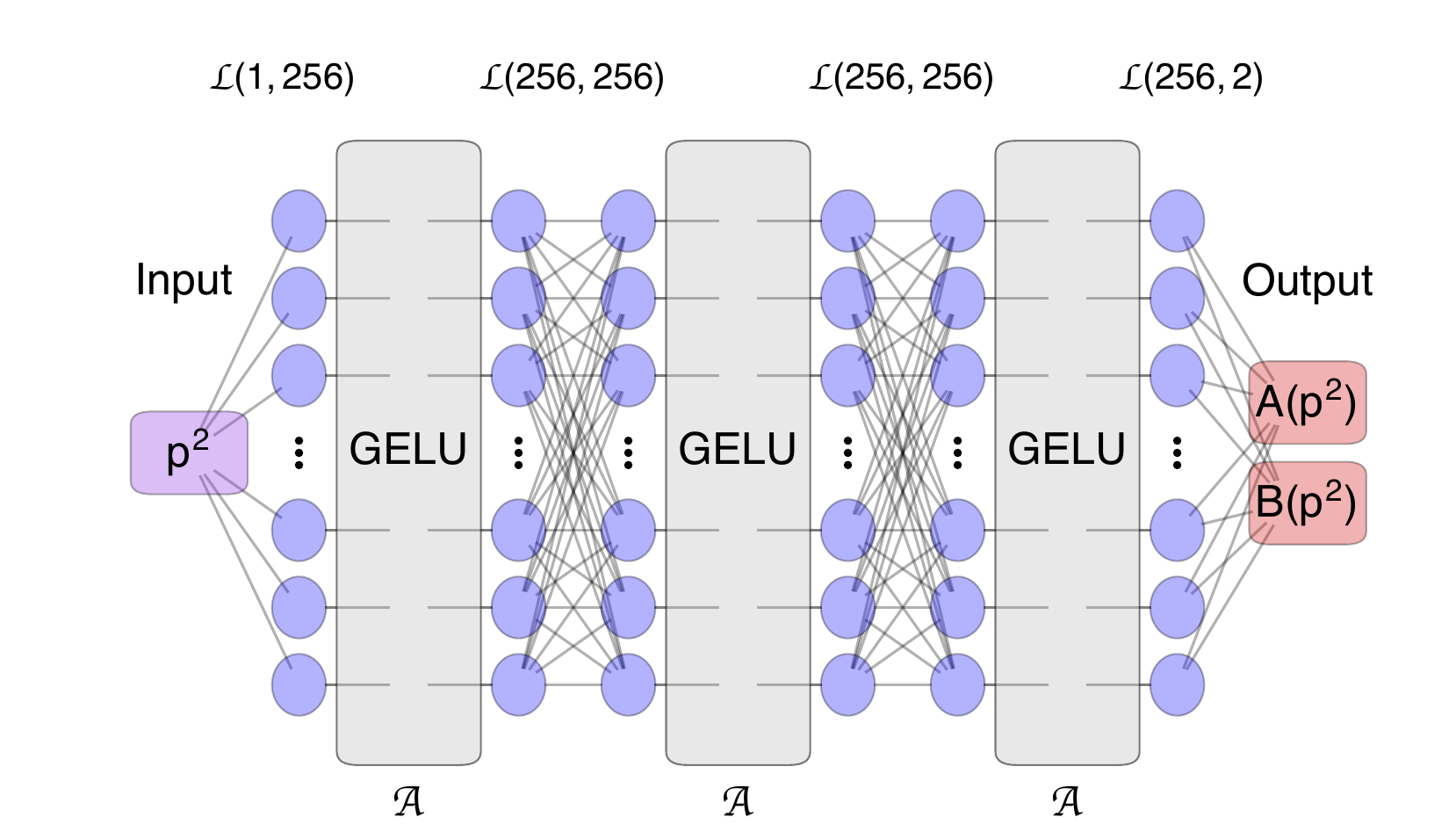}
    \put(2,55){\textsf{\large C}}
\end{overpic}
\begin{overpic}[width=\columnwidth]{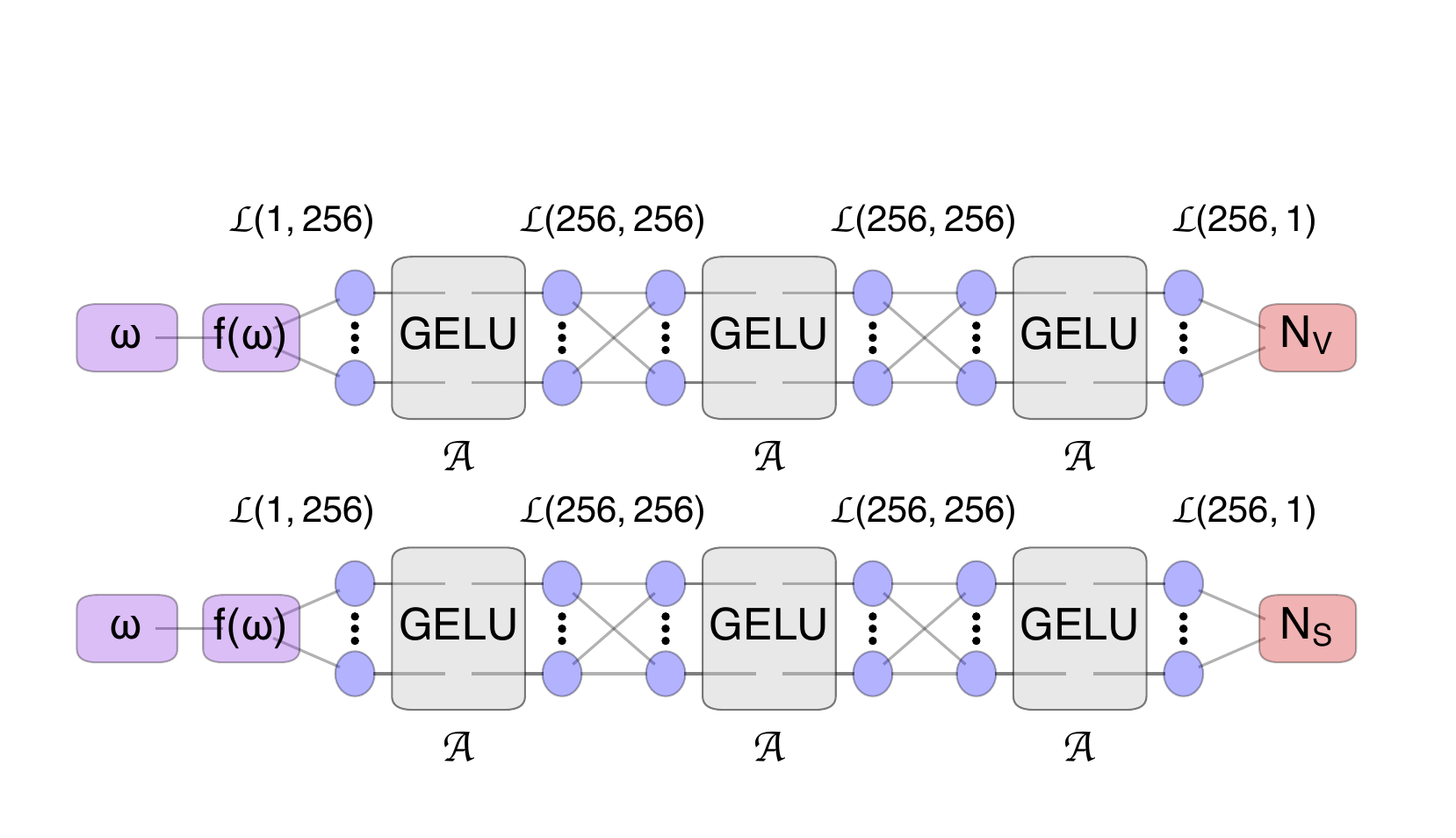}
    \put(2,55){\textsf{\large D}}
\end{overpic}
\caption{PINN architectures used in this study. \textsf{Panel A.} maps the transformed input for the solution on the real axis to $A(p^2)$ and $B(p^2)$. \textsf{Panel B.} maps the transformed radial momentum and angular variable to the real and imaginary parts of $A$ and $B$. \textsf{Panels C, D.} show the architectures used for analytic continuation and spectral reconstruction, respectively. Here, $\mathcal{L}$ and $\mathcal{A}$ denote linear and activation layers, and crossing lines indicate fully connected mappings.\label{fig:diagram}}
\end{figure*}

Gauss--Legendre quadrature is employed to generate the discrete momentum points required for numerical integration~\cite{golub1969calculation}. The squared Euclidean momentum is sampled over $p^2\in(0,\Lambda^2)$, where $\Lambda=100~\mathrm{GeV}$ is the ultraviolet regularization scale. In the architectures shown in Figs.~\ref{fig:diagram}A and \ref{fig:diagram}B, the positive radial input $x=p^2$ is mapped according to $f(x)=\ln(x/x_0)/10$, with $x_0=1~\mathrm{GeV}^2$. This dimensionless mapping compresses the wide momentum interval to an order-one numerical range and improves the conditioning of the network optimization. The mapping is not used in the analytic-continuation architecture shown in Fig.~\ref{fig:diagram}C, because the complex logarithm is multivalued and would introduce an artificial branch point and branch cut.



\subsection{\label{Hidden layer}Hidden layer}

The hidden layers consist of fully connected affine transformations followed by nonlinear activation functions~\cite{Goodfellow:2016dl}. For the networks used to solve the gap equation on the real axis and in the complex plane, all input variables are real and we employ the bounded and smooth $\tanh$ activation function to facilitate stable optimization. However, the analytic-continuation network is evaluated directly at complex values of $p^2$. Since the complex extension of $\tanh z$ is meromorphic and contains an infinite sequence of poles, its use in this network would introduce artificial singularities into the complex momentum plane.

To avoid such artificial singular structures, we instead employ the Gaussian Error Linear Unit (GELU) activation function for the analytic-continuation network,
\begin{align}
    \operatorname{GELU}(z)=
    \frac{z}{2}
    \left[
        1+\operatorname{erf}\left(\frac{z}{\sqrt{2}}\right)
    \right],
\end{align}
where
\begin{align}
    \operatorname{erf}(z)=
    \frac{2}{\sqrt{\pi}}
    \int_{0}^{z} e^{-t^2}\,dt
\end{align}
is the error function. Since $\operatorname{erf}(z)$ is an entire function, GELU is also entire and therefore introduces neither poles nor branch cuts at finite complex arguments. Moreover, GELU is smooth throughout the complex plane, which facilitates gradient-based optimization. These properties make GELU particularly suitable for constructing a neural-network representation whose analyticity in the complex momentum plane is explicitly preserved.

\subsection{Output layer\label{Output layer}}

The output layer defines the final form of the solution function output by the PINN. In numerous research scenarios~\cite{Ullah:2024bkj,Berrone:2023,Sukumar:2022,Terin:2024iyy,Aygun:2023}, explicit physical constraints can be imposed on the network output to satisfy fundamental physical properties of the system. For example, in Ref.~\cite{Terin:2024iyy}, the network output is constrained to be strictly positive using the function $\mathrm{Softplus}(x)=\ln\left(1+\exp(x)\right)$, thereby ensuring consistency with the underlying physical requirements. For the spectral reconstruction, we impose the threshold and ultraviolet constraints directly on the network outputs. To keep the neural networks as simple as possible, we use their outputs directly to represent the dressing functions when solving the gap equation on the real axis and in the complex plane. In the spectral reconstruction, however, training on a finite momentum domain makes it difficult for the networks to learn the ultraviolet asymptotic behavior accurately. We therefore introduce explicit prefactors in the output layer to enforce both the required asymptotic behavior and the spectral threshold condition. For $\omega\geq\omega_{\mathrm{th}}$, the prefactors entering the vector and scalar spectral densities are
\begin{subequations}
\begin{align}
f_{\mathrm{ir}}(\omega)
&=\frac{\omega-\omega_{\mathrm{th}}}
       {\omega-\omega_{\mathrm{th}}+\omega_{\mathrm{ir}}},\\
f_{\mathrm{uv}}^V(\omega)
&=\left(\frac{\omega_{\mathrm{uv}}}
              {\omega+\omega_{\mathrm{uv}}}\right)^2,\\
f_{\mathrm{uv}}^S(\omega)
&=-\frac{\hat{m}\gamma_m\pi}
        {2(\omega+\omega_{\mathrm{ir}})
         [L_\omega'(\omega)]^{1+\gamma_m}},\\
f_{\mathrm{match}}(\omega)
&=\frac{\omega_{\mathrm{match}}}
       {\omega+\omega_{\mathrm{match}}},
\end{align}
\end{subequations}
where $L_\omega'(\omega)=\tfrac{1}{2}\ln\!\left(e^2+\omega/\Lambda_{\mathrm{QCD}}^2\right)$.
We use $\omega_{\mathrm{ir}}=2.0\times10^{-3}~\mathrm{GeV}^2$,
$\omega_{\mathrm{uv}}=3.0~\mathrm{GeV}^2$, and
$\omega_{\mathrm{match}}=0.7~\mathrm{GeV}^2$.

To select a particular solution branch of the quark gap equation, a parameter seeding strategy can be adopted. Specifically, we first conduct a small number of training iterations in the neighborhood of the target solution to obtain an approximate solution. Subsequently, the obtained preliminary parameters are set as the initialization seed for formal training, and high-precision solutions can be obtained through further optimization. This seeding method can efficiently capture the required specific solution without incurring significant additional computational cost~\cite{Guo:2023,Huang:2022}.

\subsection{Training set, test set, and loss function\label{Training set}}

The training set contains the collocation points used to minimize the DSE residual, whereas the independent test set provides additional momentum points at which the residual is evaluated. Comparing the training and test losses provides a diagnostic of generalization and helps identify the onset of overfitting~\cite{Hastie:2009itz}. This distinction is particularly important in physics-informed learning, where the objective is to approximate a solution of the governing equations rather than merely interpolate sampled data~\cite{Raissi:2017zsi}.

As we illustrate in Appendix~\ref{Input layer}, both datasets require discrete sampling within the computational domain. To facilitate high-precision numerical integration in subsequent calculations, all discrete points are generated by Gauss--Legendre quadrature. Because the dressing functions vary rapidly in the infrared region, a denser sampling of the small-momentum domain is used to improve the accuracy of the numerical integration. We generate the test grid on the same momentum domain using a different Gauss--Legendre quadrature order, so that its nodes are distinct from the training nodes. Because the Gauss--Legendre nodes are nonuniformly distributed, all mean relative errors quoted in the main text are evaluated as normalized domain-measure-weighted averages rather than simple arithmetic means over the sampled points.

For the gap equation solvers, the loss function measures the residual of the quark gap equation. Throughout this work, the individual loss terms are constructed using mean squared errors (MSEs). The mathematical simplicity and differentiability of the MSE make it well suited to gradient-based network optimization~\cite{Goodfellow:2016dl,bishop2006pattern}.

To obtain the solutions on the real momentum axis and in the complex momentum plane, we train the corresponding networks by minimizing the residual of the gap equation. Let $X\in\{A,B\}$, and define
\begin{align}
    \mathcal{R}_{X}(p_i^2;\theta)=
    \mathcal{H}_{X}[A_\theta,B_\theta](p_i^2)-X_\theta(p_i^2),
\end{align}
where $\mathcal{H}_{X}$ denotes the corresponding right-hand side of the DSE, and $\theta$ denotes the neural-network parameters. Since the two dressing functions have different characteristic magnitudes, we introduce a separate normalization scale for each component:
\begin{align}
    s_X
    =
    \sum_{i=1}^{N}
    \left|
    \mathcal{H}_{X}[A_\theta,B_\theta](p_i^2)
    \right|^2,
\end{align}
where $N$ is the number of training collocation points. The DSE loss is then
\begin{align}
    \mathcal{L}_{\mathrm{DSE}}
    =
    \sum_{X\in\{A,B\}}
    \frac{1}{s_X}
    \sum_{i=1}^{N}
    \left|\mathcal{R}_{X}(p_i^2;\theta)\right|^2.
    \label{eq:dse-loss}
\end{align}
For the solution in the complex momentum plane, the squared modulus in Eq.~(\ref{eq:dse-loss}) includes both the real and imaginary parts of the residual. The network used for analytic continuation is trained with the same residual evaluated only at collocation points on the real momentum axis.

The spectral reconstruction is an inverse problem in which the continuous spectral densities and pole parameters are inferred from the propagator. The vector and scalar components differ in magnitude and vary on different momentum scales. A single global MSE can therefore give too little weight to the smaller scalar component or to narrow regions of rapid variation. We therefore employ the composite objective $\mathcal{L}_{\mathrm{spec}}=\sum_k\lambda_k\mathcal{L}_k$, where $\mathcal{L}_k$ denotes an individual MSE-based constraint and $\lambda_k$ its weight. The objective combines propagator reconstruction, dressing-function matching, region-dependent absolute and relative errors, derivative matching, and spectral smoothness.

Both objectives are minimized using the Adam optimizer. For the gap equation solvers, a \texttt{ReduceLROnPlateau} scheduler halves the learning rate when the training loss plateaus, allowing the network parameters to be automatically fine-tuned.
\begin{acknowledgments}
We are grateful to Craig Roberts, Daniele Binosi, and M.-H.~Ding for their guidance and insightful discussions throughout this work. We thank Z.-Q.~Yao, Z.-N.~Xu, K.-X.~Zeng, K.-Y.~Guo, Y.-Y.~Xiao, X.-Y.~Bai, W.-H.~Bian, and other colleagues for valuable discussions.
This work was supported by
National Natural Science Foundation of
China, grant no.\ 12135007 and
the Program of China Scholarship Council, grant no.\ 202606190076.
\end{acknowledgments}
\bibliography{PINN}
\end{document}